\documentclass[a4paper,fleqn,usenatbib]{mnras}

\usepackage{newtxtext,newtxmath}

\usepackage[T1]{fontenc}
\usepackage{ae,aecompl}

\usepackage{graphicx}	
\usepackage{amsmath}	
\usepackage{amssymb}	
\usepackage{epstopdf}
\usepackage{natbib}
\usepackage{longtable}
\usepackage{lscape}
 \usepackage{tabu}
 \usepackage{multirow}
\usepackage{tabu}
\usepackage{widetext}
\usepackage{supertabular}       
\usepackage{longtable}

\graphicspath{{./Figures/}} 

\title[Photometric rotation periods for 107 M dwarfs]{Photometric rotation periods for 107 M dwarfs from the APACHE survey}

\author[P. Giacobbe et al.]{P. Giacobbe,$^{1}$\thanks{E-mail: paolo.giacobbe@inaf.it}
M. Benedetto,$^{2}$ 
M. Damasso,$^{1}$ 
A. Sozzetti,$^{1}$ 
J.M. Christille,$^{2}$
\newauthor
M.G. Lattanzi,$^{1}$ 
P. Calcidese, $^{2}$ 
A. Carbognani, $^{5}$ 
D. Barbato,$^{4,1}$ 
M. Pinamonti,$^{1}$ 
E. Poggio, $^{1}$ 
\newauthor
A. F. Lanza,$^{3}$ 
A. Bernagozzi,$^{2}$   
D. Cenadelli, $^{2}$ 
L. Lanteri,$^{1}$ 
E. Bertolini$^{2}$ 
\\
$^{1}$Astrophysical Observatory of Torino, INAF, Strada Osservatorio 20, Pino Torinese (To) 10025, Italy\\
$^{2}$Astronomical Observatory of the Autonomous Region of the Aosta Valley, Frazione Lignan 39, Nus (Ao), 11020, Italy\\
$^{3}$Astrophysical Observatory of Catania, INAF, Via S. Sofia 78, Catania, 95123,  Italy\\
$^{4}$Department of Physics, University of Torino, Via Pietro Giuria 1, I-10125 Torino, Italy\\
$^{5}$Astrophysical and Space Science Observatory of Bologna, INAF, Via Gobetti 93/3, Bologna, 40129,  Italy\\
}

\date{Accepted XXX. Received YYY; in original form ZZZ}

\pubyear{2019}

\begin{document}
\label{firstpage}
\pagerange{\pageref{firstpage}--\pageref{lastpage}}
\maketitle

\begin{abstract}
We present rotation period measurements for 107 M dwarfs in the mass range $0.15-0.70 M_\odot$ observed within the context of the APACHE photometric survey. We measure rotation periods in the range 0.5-190 days, with the distribution peaking at $\sim$ 30 days.
We revise the stellar masses and radii for our sample of rotators by exploiting the \emph{Gaia} DR2 data.
For $\sim 20\%$ of the sample, we compare the photometric rotation periods with those derived from different spectroscopic indicators, finding good correspondence in most cases. We compare our rotation periods distribution to the one obtained by the \textit{Kepler} survey in the same mass range, and to that derived by the MEarth survey for stars in the mass range $0.07-0.25 M_\odot$. The APACHE and \textit{Kepler} periods distributions are in good agreement, confirming the reliability of our results, while the APACHE distribution is consistent with the MEarth result only for the older/slow rotators, and in the overlapping mass range of the two surveys. 
Combining the APACHE/\textit{Kepler} distribution with the MEarth distribution, we highlight that the rotation period increases with decreasing stellar mass, in agreement with previous work. Our findings also suggest that the spin-down time scale, from fast to slow rotators, changes crossing the fully convective limit at $\approx0.3 M_\odot$ for M dwarfs.
The catalogue of 107 rotating M dwarfs presented here is particularly timely, as the stars are prime targets for the potential identification of transiting small planets with TESS and amenable to high-precision mass determination and further atmospheric characterization measurements.
\end{abstract}




\section{Introduction}

The efficacy of the two techniques (transit photometry and Doppler spectroscopy) that have enabled the detection of the overwhelming majority of exoplanets known to-date ($\sim 4000$, see e.g. \url{https://exoplanetarchive.ipac.caltech.edu/} and \url{http://exoplanet.eu/}) can be fundamentally hampered by insufficient understanding of photometric and spectroscopic signals that are stellar and not planetary in nature. For instance, stellar magnetic activity cycles with typical timescales of several years of duration can be a serious concern when claiming detection of long-period planets in radial-velocity (RV) time-series (e.g., \citealt{carolo14,endl16}). In such cases, the ability to discriminate the true nature of the signals crucially depends on the availability of spectroscopic proxies for stellar activity and/or photometric monitoring. When it comes to M dwarf type planet hosts, the stellar temperate zone, within which the signals of potentially habitable planets can today be readily identified in principle (e.g., \citealt{anglada16,dittmann17,mascareno17,astudillo17b,bonfils18,zechmeister19}), corresponds to orbital periods typically in the range of tens of days (e.g., \citealt{kopparpu13,kopparapu14,kopparapu18}). However, these timescales can coincide with those of stellar rotation periods of low-mass stars in the middle of their main-sequence lifetime (e.g., \citealt{barnes07,mcquillan13,vanderbourg16,newton2016,suarezmascarenoetal2018}). Without a detailed understanding of stellar activity-induced effects in RVs to supporting spectroscopic and photometric information, ambiguities in the interpretation of RV signals with periods corresponding to habitable zone distances might be long-lasting (e.g., \citealt{robertson14,anglada15,robertson15,anglada16b}). Even in the case of transiting systems, for which the periodicity of the planetary signal can be unambiguously identified via transit photometry, the proximity between orbital and insufficiently well characterized rotation periods can undermine our ability to determine dynamical masses with high statistical confidence (e.g., \citealt{cloutier17,damasso18,damasso19}).

The fraction of low-mass M dwarfs harbouring temperate super Earth-type planets is likely to be high (e.g., \citealt{bonfils13,dressing15,hardegree19}). This is one of the reasons why they figure prominently as main targets of important ground-based RV surveys (e.g., \citealt{afferetal2016,astudillo17c,luque18,hobson19}). For the Transiting Exoplanet Survey Satellite (TESS; \citealt{ricker14}), launched in 2018 April, small habitable-zone planets around M dwarfs constitute a particularly relevant sample (e.g., \citealt{sullivan15,barclay18,ballard19,kaltanegger19}), as the combination of small telescope aperture and satellite's observing strategy translate in a low mission sensitivity to orbital periods longer than a few tens of days (except near the ecliptic poles) and to planets with radii $\lesssim2$ R$_\oplus$ around solar-type primaries. Measurements of photometric rotation periods for M dwarfs with TESS-detected small-size transit candidates would be of paramount importance to gauge the effects of spots in the analysis of the transit events and to help disentangling planetary RV signals from those related to stellar magnetic activity (e.g., \citealt{haywood14,vanderbourg16, dittmann17,damasso18}). However, rotation periods of a few tens of days or longer will not be typically accessible to TESS (at least during its nominal mission lifetime), due to its less than a month duration of the observing windows. Follow-up efforts can thus take advantage from the availability of additional data that can aid in the characterization of the detected systems.

In this paper we present rotation periods for 107 nearby, early- to mid-M dwarfs in the northern hemisphere using photometric time-series with a multi-year time baseline gathered within the context of the APACHE (A PAthway towards the CHaracterization of Habitable Earths) photometric transit search project \citep{sozzetti13apa}. With TESS now beginning to survey the northern hemisphere, the study presented here is particularly timely, as all of the bright M dwarfs in the sample objective of this work are prime targets for the potential identification of transiting small planets amenable to high-precision mass determination and further atmospheric characterization measurements (TESS mission Level 1 Requirement). Our sample of rotation periods can also be used to further our understanding of important issues of stellar astrophysics pertaining to partly and fully convective stars, such as differences in rotational evolution (e.g., \citealt{gilhool18}, and references therein) and the age-rotation-activity relation (e.g., \citealt{wright18,gonzalez19}).

The paper is organized as follows: in Sec.\,\ref{secapa} we introduce the APACHE photometric survey; in Sec.\,\ref{secinput} we describe our input catalogue; in Sec.\,\ref{results} we present our results; in Sec.\,\ref{secwide} we compare our findings with the literature from the spectroscopic survey HADES and the photometric surveys \textit{Kepler} and MEarth; in Sec.\,\ref{sum} we summarize our findings and we discuss some points of interest related to the exoplanet research and the M dwarfs stellar physics.

\section{The APACHE survey}
\label{secapa}
\subsection{Infrastructure}
\label{apasurv}

APACHE employs an array of five 40-cm telescopes hosted on a single platform with a roll-off enclosure, located at the Astronomical Observatory of the Aosta Valley (OAVdA), in the western Italians Alps, at 1650 meters above the sea level. The site characterization study was presented in \cite{damasso10}, while a feasibility study of the APACHE project was presented in \cite{giacobbe12}.
The telescope array is composed of five identical Carbon Truss 40-cm f/8.4 Ritchey-Chr{\'e}tien telescopes, with a GM2000 10-MICRON mount and equipped with a FLI Proline PL1001E-2 CCD Camera and Johnson-Cousins V \& I filters.
The pixel scale is $1.5"/pixel$, yielding a field-of-view (FOV) of $21'$x$21'$.
The open source observatory manager RTS2 \citep{kubanek10} was the choice for the high-level software
control of the five-telescope system, including dynamic scheduling of the observations \citep{christille13}.

\subsection{Observational strategy}

As APACHE is a targeted survey, its sampling strategy differs from that of  wide-field transit surveys (e.g., HAT-Net; \citealt{bakos18} , Super-WASP;\citealt{pollacco06}), and it more closely resembles that of other experiments, which adopted the 'one target per field' approach (e.g., MEarth and MEarth-South; \citealt{nutzman08}).
The optimal observing strategy should maximize the number of targets observed per night while preserving a time sampling good enough to detect transit events due to short-period planets.
Exploiting the data collected within the pilot study \citep{giacobbe12}, a series of detailed simulations with different temporal sampling and number of exposures was performed in order to choose the optimal observing strategy. In particular, we tested intervals of 10 through 50 minutes between two sets of consecutive pointings of the telescope on the same target, and 1 to 5 consecutive exposures for each pointing. Eventually, we adopted for APACHE an observing strategy consisting of 3 consecutive exposures every 20 minutes.
In this way, during a typical night of observation, each telescope observes $\sim 12$ fields, where the grater part of them contain only a single target M dwarf.
Each target is observed for the whole time available during the night with airmass below 2.
Exposure times are selected to yield a signal to noise ratio (SNR) for the target star $>$ 200 while avoiding detector saturation.

\subsection{Data reduction and differential photometry analysis pipeline}
\label{diff_phot}
Data reduction and analysis were carried out using a dedicated software package written in IDL, which utilizes free libraries from the Astronomy User's Library and external routines in FORTRAN and C++. It is organized into modules:
\begin{itemize}
\item image calibrations (dark, bias and flat-fielding subtraction);
\item image alignment (via astrometric solution) and photometric processing (aperture photometry);
\item differential photometry and trend filtering.
\end{itemize}
While image calibration and alignment implement standard procedures for which it is not necessary to provide lengthy descriptions, the extraction of the light curve in the third module of the pipeline deserves more attention, so we describe it here in more details. This component of the pipeline performs those operations that are necessary to correct, to a high degree of reliability, for systematic effects that cause the degradation of the photometric quality and consequently of the transit detection efficiency. 
This is a fundamental step of the pipeline because it provides the starting point, but also the validation benchmark, for more sophisticated filtering procedures.

We start by applying a straightforward differential photometry technique: for each frame $i$, we use the average magnitude of $n$ reference stars $M_{ref}(i)$ according to the equation
\begin{equation}
   M_{ref}(i)=\frac{\sum_{k=0}^n M_k(i)}{n} 
\end{equation}
where the $M_k$ time-series are zero-averaged.
The normalized magnitude of the target $M_{target}(i)$ is then subtracted from $M_{ref}(i)$ , obtaining the difference $\Delta M(i)$
\begin{equation}
    \Delta M(i) = M_{ref}(i) - M_{target}(i)
\end{equation}
corrected for all the common systematics.
In this process there are two key points: i) we need the most accurate estimate of the instrumental magnitude $M_{ref}(i)$ and $M_{target}(i)$ and ii) we need the best set of reference stars (e.g excluding variable stars or stars affected by peculiar systematic errors like bad pixels).
For the point i), we have implemented a multi-aperture photometric processing. We settled on 12 apertures, typically ranging from two to four times the average full width at half-maximum (FWHM) of the point spread function.
While for the point ii), we first take care of picking up reference objects on a CCD subframe, avoiding the chip edges, affected by vignetting which is not fully corrected for during flat-fielding. Secondly, in order to choose the appropriate set of references for the target we use a method based on the \cite{burke06} prescription. This method selects the subset of reference stars which minimizes the RMS of the differential light curve of the target and it is then applied to all 12 apertures in order to choose the optimal one, on the basis of a minimum-RMS prescription for the target light curve. The target and the reference stars use the same selected aperture.

Although the standard photometric procedure performs very well on nightly basis, there are some residual systematic effects after this processing.
As described before, APACHE uses German Equatorial Mounts, which necessitate effectively rotating the telescope through $180^{\circ}$ relative to the sky when crossing the meridian. Thus, each set of reference stars (not the target that is always in the center of the detector) falls on two areas of the CCD, one for negative and one for positive hour angle.
In this situation, flat fielding errors manifest themselves as different base-line magnitudes on each side of the meridian. Following the discussion in \cite{berta12}, dealing with the same problem in MEarth, we call this effect ``mount-flip''.
Secondly, we observe correlations between the measured differential magnitudes of the target M dwarfs and weather parameters, specifically the sky brightness (which depends in turn from the presence of clouds/cirri and the lunar phase).
Thirdly, we observe a correlation between the amplitudes of the systematics that affect the targets and the amplitudes of the (same) systematics that affect the reference stars.  
Probably, the last two correlations are produced by a mismatch, in particular in spectral type, between the target star and the comparison stars. In other words, each target over a period of years shares a linear combinations of the systematics of the field. Considering this, we can't guarantee with the ``standard'' procedure the stability of the systems at mmag level over a period of many years.

Furthermore, we investigate the effect of the precipitable water vapour, a systematic effect with a significant impact on the photometric performance of other M dwarfs surveys, such as MEarth \citep{berta12}.
We follow closely the approach proposed by \cite{berta12}, looking for a ``common mode'' across our M dwarfs sample, without finding a clear correlation with it and, consequently, with no need to build a common mode function. This is likely due to two reasons: 1) we observe with the $I$-band filter, a photometric band less affected by telluric absorption with respect to the redder custom-made MEarth filter; 2) our sample is mainly constituted by early-type M dwarfs with a color index more similar to that of field stars.

In this context, a reliable filtering algorithm is mandatory to reach the photon noise level over the full observation period. 
Our best solution adapts the Trend Filtering Algorithm (TFA, \citealt{kovacs05}) to our purpose and peculiarity. 
The principal goal of our trend filtering algorithm is to create a so called ``filter function'' representing the systematic effects influencing the target light curve. This ``filter function'' can be considered as a data-driven model where there is no a priori knowledge of the systematics that we want to filter and it is built directly from the field stars light curve. 

The first step is to collect the reference stars light curves in a matrix $X$.
To build $X$ we consider the full set of reference stars excluding only those stars clearly variable, on the basis of the deviation between each star's RMS light curve derived by the standard photometry and the theoretical single point uncertainty derived by Eq.\,\ref{eq1}.
While for the differential photometry we generally select a set of few reference stars, our trend filtering algorithm requires as many reference stars as possible, in order to be sure to include all the systematics of the field. 
Furthermore, as mentioned before, there are a set of systematics poorly sampled by the reference stars but strongly present in the final differential light curves.
We insert them directly inside the $X$ matrix.
To solve the ``mount-flip'' we create a single-row matrix that models the orientation of the mount (we put a value of 0 if the mount was oriented to East, 1 otherwise) and we add it as a new row of the $X$ matrix.
Similarly, we create a model representing the ``global'' sky background (inserting frame by frame its measured mean value on the field) and we add it as a new row of the $X$ matrix as explained before.
The sky background for each of the stars in the FOV is computed using an annulus around the source with dimensions generally between four and seven times the aperture for the sources' photometry.

Then we create a model of the filtered light curve $A(i)$
\begin{equation}
    A(i) = \displaystyle\sum_{j=0}^M \frac{X_j(i)}{M}
\end{equation}
where M is the total number of reference objects plus ``mount-flip'' and sky background.
The filter function $F(i)$ is now defined according with the equation
\begin{equation}
    F(i)= \displaystyle\sum_{j=0}^M c^j X^j(i)
\end{equation}
To determine the $c^j$ coefficient we minimize the expression $D$ 
\begin{equation}
    D= \displaystyle\sum_{i=0}^N [Y(i)-A(i)-F(i)]^2   
\end{equation}
via Singular Value Decomposition (SVD), where $Y$ is the target light curve vector and $N$ is the number of observations.	
The filtered light curve $\tilde{Y}(i)$ is finally obtained by 
\begin{equation}
  \tilde{Y}(i)= Y(i)-F(i)  
\end{equation}

\begin{figure}
    \includegraphics[width=1\linewidth]{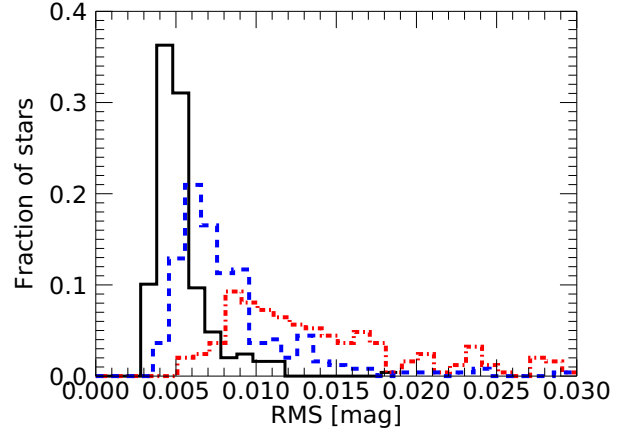}
    \caption{Fraction of stars as a function of RMS for the APACHE sample. The black solid line shows the distribution of the single point uncertainty $\sigma_t$, the red dotted line shows the distribution of the RMS precision of the ``standard'' differential photometry, the blue dashed line shows the distribution of the RMS precision of the light curves detrended with our trend filtering algorithm.}
    \label{fig1bis}
\end{figure}

We consider the RMS of the light curves as a proxy of the impact of this kind of systematic correction over the data quality.
In Fig.\,\ref{fig1bis} we show the distribution of the single point uncertainty $\sigma_t$ (black solid line), the RMS precision of the ``standard'' differential photometry (red dotted line) and the RMS precision of the light curves detrended with our trend filtering algorithm (blue dashed lines).
While the RMS precision of the ``standard'' differential photometry is heavily affected by systematic effects, with our filtering we are able to approach the theoretical distribution with a clear improvement of the data quality. 

Sometimes this kind of correction could be too aggressive and it might contribute to suppress the signal amplitude. 
As described in Sec.\,\ref{period_d}, after the steps just mentioned we use the Generalized Lomb-Scargles (GLS) algorithm \citep{Zechmeister09} to find a sinusoidal signal, with periodicity and phase. If the signal has a False Alarm Probability (FAP) $\leq$ 1\%,  we use the \cite{foreman15} approach to preserve and rebuild it.  With the GLS results we create a rotation model with amplitude equal to 1, then we add it to the X matrix as a new row. Applying TFA again with the new X matrix we treat the signal as a systematic, we correct it and we obtain its amplitude. We therefore have a completely corrected light curve, the signal period, phase and amplitude. Once rebuilt, the rotation signal is added to the reduced light curve.
A similar approach, with the goal to filter out the systematics without affecting the signal, was developed by \cite{berta12}. The main, but substantial difference is that the algorithm by \cite{berta12} performs the filtering plus signal research after the standard differential photometry (that is a sort of pre-whitening) while our method, based on the framework established by \cite{foreman15}, performs the filtering  plus signal search together with the differential photometry.
Nevertheless, the idea to describe the systematics poorly sampled by the reference stars using some analytical functions closely follow \cite{berta12}.
The combination of the two techniques ultimately allows for a more thorough treatment of systematics.

\section{The input catalogue}
\label{secinput}

\subsection{Initial target selection}
\label{aic}
The APACHE Input Catalogue was built as a sub-list of the all-sky sample of 8889 bright (magnitude $J < 10$) low-mass stars in \cite{lepine11}. After checking the visibility constraints from OAVdA (at least 3 hours per night with altitude $\geq 30^{\circ} $ and over a period of at least 2 months), the number of potentially good (at least 5 stars with V magnitude within 1 magnitude from the target) comparison stars in the telescopes' field-of-view and the absence of relevant blended objects, we selected $\sim$3000 targets composing our final Input Catalogue.
Then, we devised a ranking system to identify targets with higher priorities taking into account the survey scientific goals (in particular the detection of small size planets) and architecture.
The ranking was principally based on the best observability during the year and on V magnitude of the targets, considering it as a key point for the RV followup.
Then, we consider the number of the \emph{Gaia} transits based on accurate representation of \emph{Gaia}'s scanning law, in order to prioritize those in areas of the sky with higher numbers of \emph{Gaia} astrometric measurements.
Furthermore, we cross-correlate with approximately two dozen catalogs, searching for additional and more precise information than that included in \cite{lepine11}, such as \textit{i)} a better determination of the spectral class, to avoid spectral types different from M dwarfs; \textit{ii)} measurements of projected rotational velocity \textit{V$\sin$i}, to favour slow rotators in that they are more suitable for precise RV follow-up; \textit{iii)} level of chromospheric activity and X-ray emission, to flag active targets which are not optimal to search for low-mass planets with the RV technique.
Given the huge impact for the exoplanet research of the synergies between spectroscopic and photometric measurements, we override our prioritization for all that targets selected for high-precision RV monitoring with the HARPS-N spectrograph on the Telescopio Nazionale Galileo (TNG) within the Global Architecture of Planetary Systems (GAPS) (e.g. \citealt{desidera14,benatti16}) and HARPS-N red Dwarf Exoplanet Survey (HADES) collaborations (e.g. \citealt{afferetal2016,pergeretal2017,affer19}).

\subsection{Observations logbook and performances}

\begin{figure*}
$\begin{array}{cc}
\includegraphics[width=0.45\linewidth]{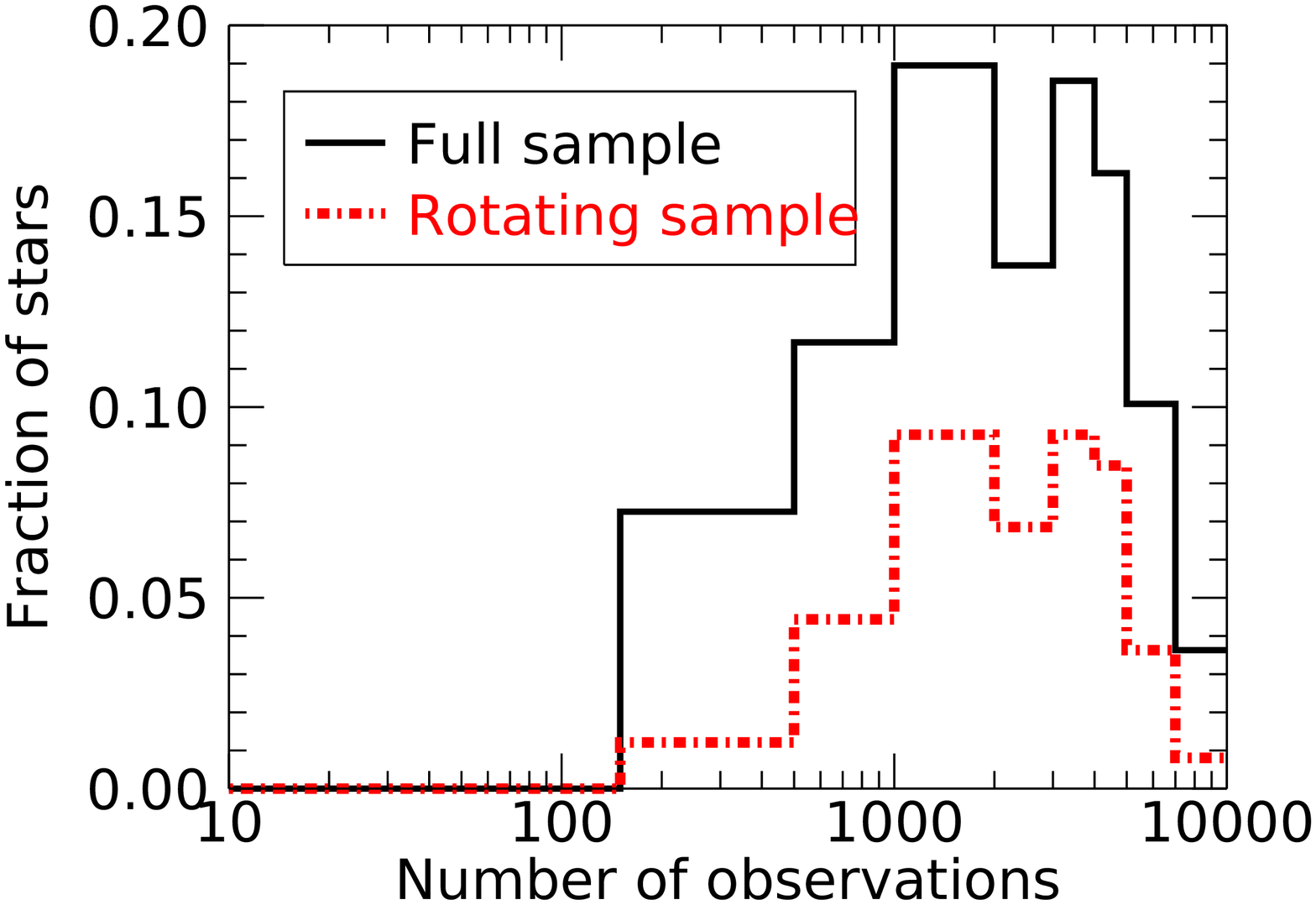} &
\includegraphics[width=0.45\linewidth]{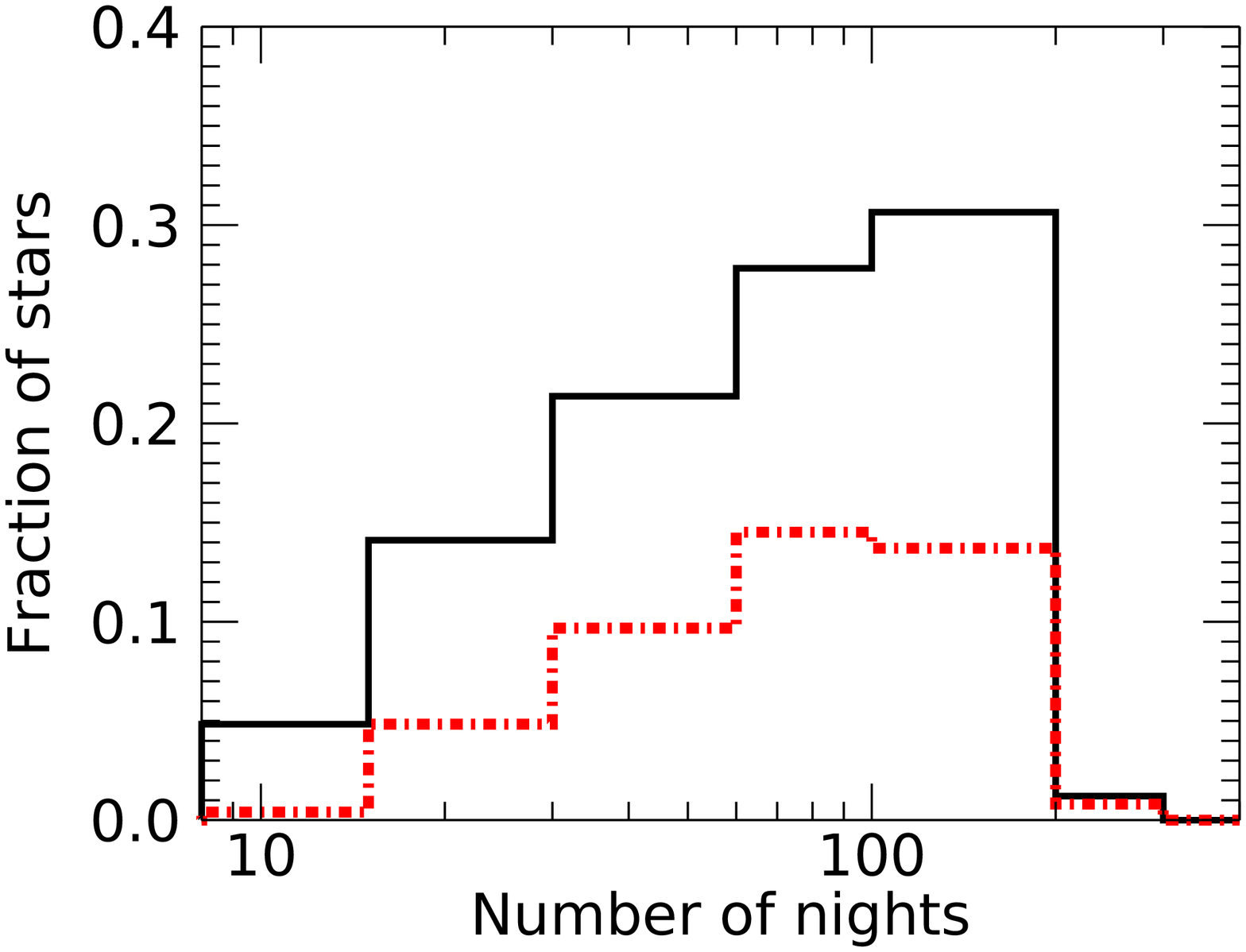} \\
\includegraphics[width=0.45\linewidth]{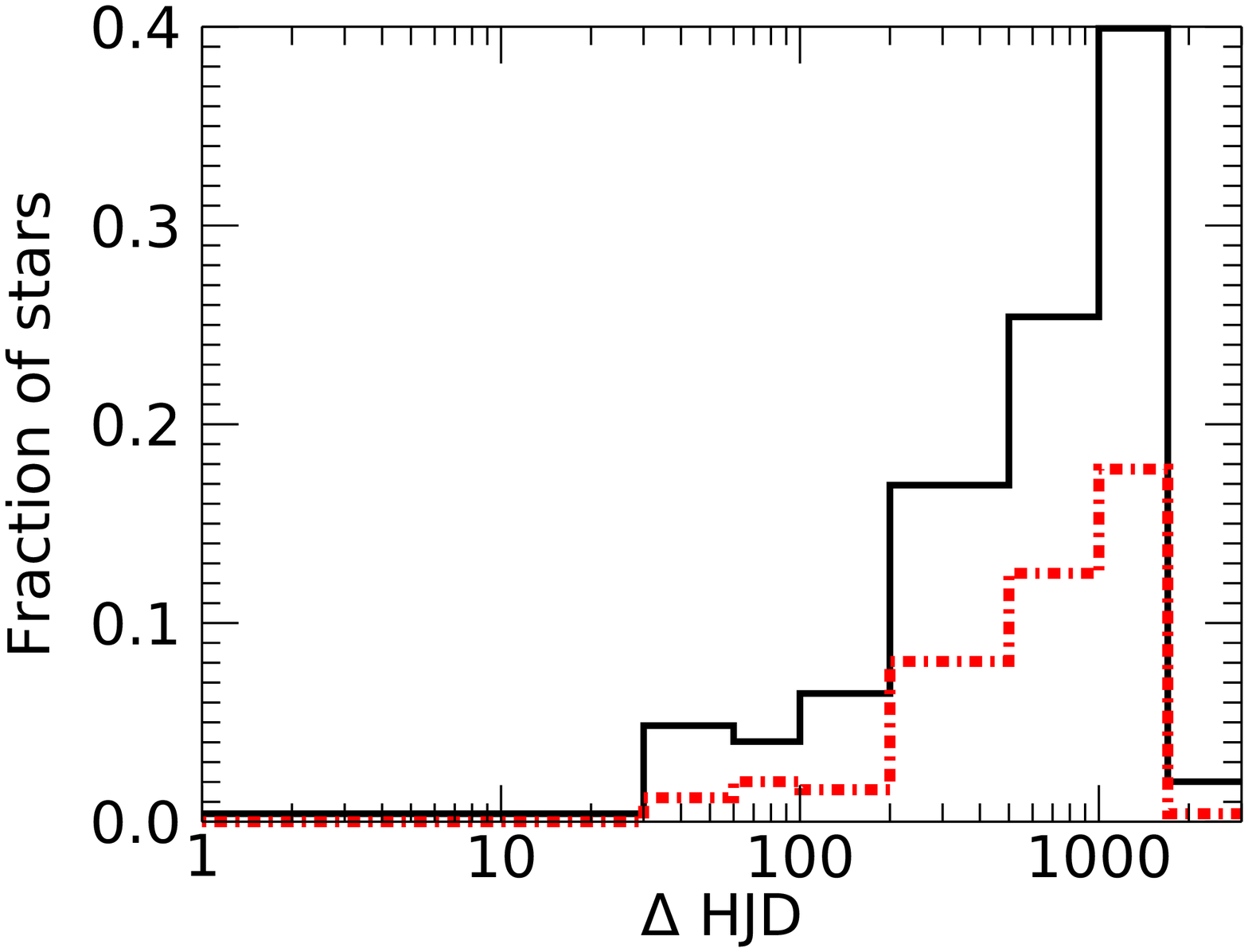} &
\includegraphics[width=0.45\linewidth]{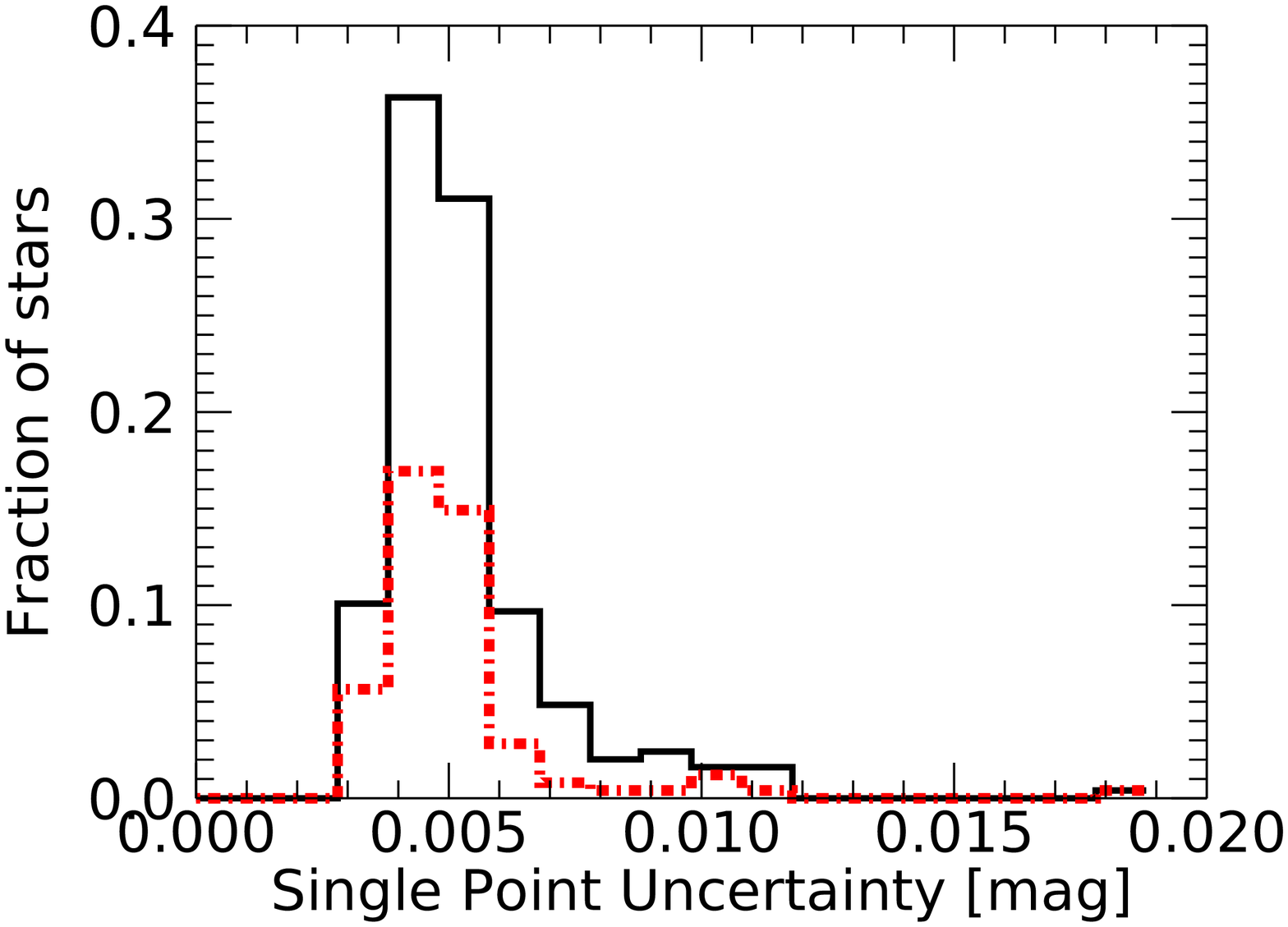} \\
\end{array} $
 \caption{Top left: fraction of stars as a function of the number of observations. Top right: fraction of stars as a function of the observation nights.
 Bottom left: fraction of stars as a function of the time span between the first and the last observation. Bottom right: fraction of stars as a function of the mean theoretical single point uncertainty. The black solid line represents the full sample of 247 M dwarfs considered for the rotation period research while the dash dotted red line represents the sample of the sample of 107 M dwarfs with detected rotational modulation.}
\label{fig1}
\end{figure*}

The data considered in this paper are the results of five year of observation between  9th July 2012 and 9th July 2017. This period corresponds to the nominal duration of the survey.
From the APACHE observations database, we select a sub-sample of 247 M dwarfs with more than 200 data points taken on at least 10 observation nights and spanning at least 30 days.
We obtained a typical per measurement theoretical single point uncertainty of $\sim$ 0.005 mag (see the bottom right panel in Fig.\,\ref{fig1}) and a median long-term RMS precision on the light curves of $\sim$ 0.007 mag (see Sec.\,\ref{diff_phot}  for further details and Fig.\,\ref{fig1bis}).

The theoretical single point uncertainty $\sigma_t$, in magnitudes, is defined by 
\begin{equation}
\label{eq1}
  \sigma_t = \frac{2.5}{ln 10} \times \frac{\sqrt{N_{star}+N_{sky}+\sigma^2_{scint}}}{N_{star}}   
\end{equation}
where $N_{star}$ is the number of photons from the source, $N_{sky}$ is the number of photons from the sky background for the photometric aperture that includes read and dark noise and $\sigma_{scint}$ is the scintillation noise (\citealt{young67}).

Fig.\,\ref{fig1} summarizes the overall properties of the 247 selected M dwarfs. In Tab.\,\ref{tab:par2} we present the observational properties for all the targets\footnote{the light curves of all the 247 stars are available upon request by e-mailing the author}, while our results are presented in Tab.\,\ref{tab:par1} for our candidate rotators.  
Overall, 81\%  of the targets have more than 500 observations and they were observed for more than one observing season, while $\approx7$\% of the targets, considered as schedule fillers, have a low number of observations ($\leq 150$). 
Although our strategy was mainly driven by the transit search and it was not optimized for the rotation period detection, we note that the data quality and the phase coverage are very suitable to look for rotational period. We quantitatively discuss this point in Sec.\,\ref{results}.

\subsection{Revised stellar properties from \emph{Gaia} DR2}
\label{sample}
\label{mas}
\begin{figure}
    \includegraphics[width=1\linewidth]{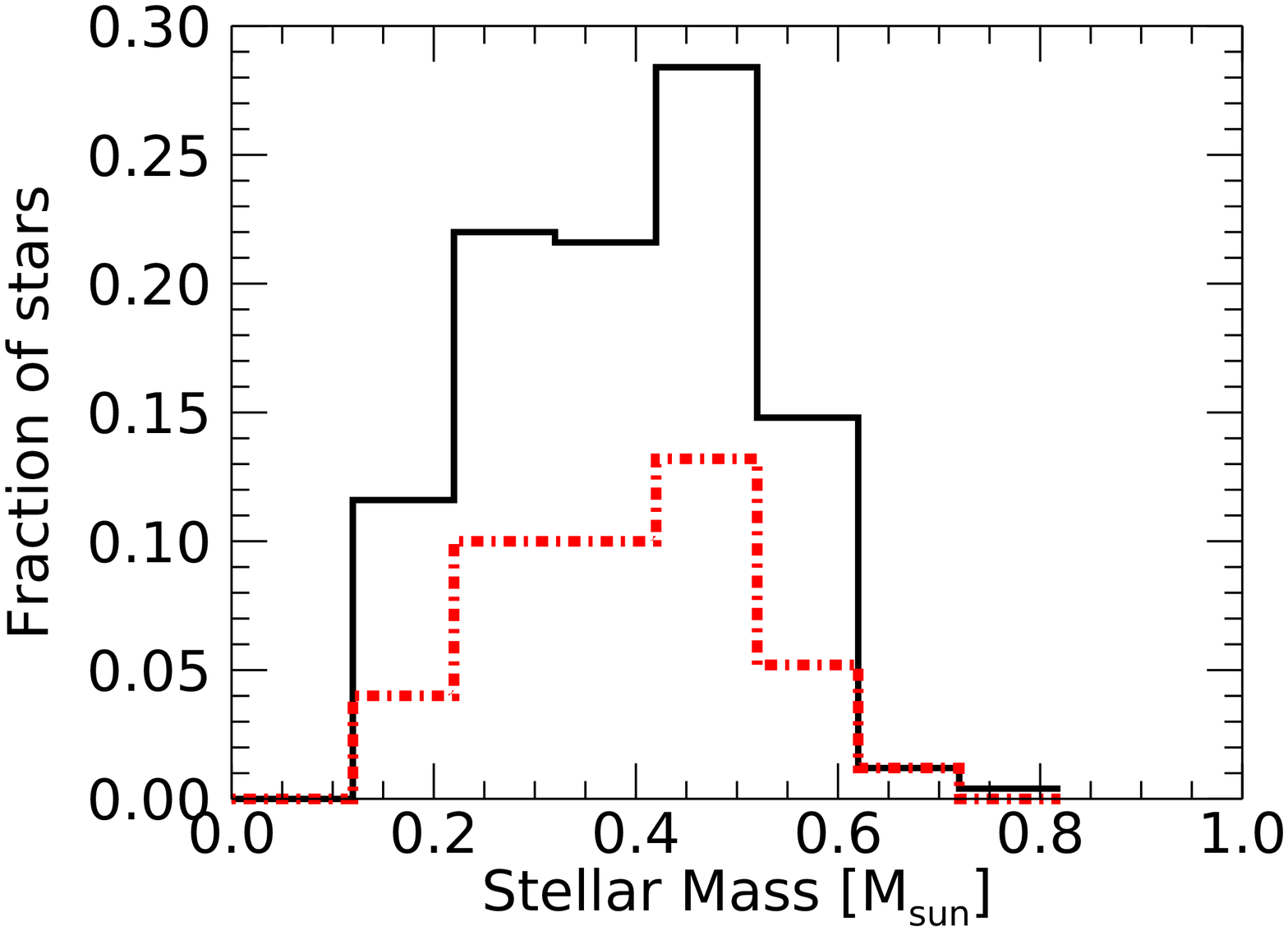}
    \caption{Fraction of stars as a function of stellar mass for the APACHE sample. The black solid line shows the distribution of the sample of 247 M dwarfs stars on which the period search was performed while the dash-dotted red line represents the distribution of the subsample with detected rotational modulation.}
    \label{fig2}
\end{figure}

For the aim of this work, it is important to assign a reliable mass and radius estimation of our M dwarf sample.
We estimate them using the \texttt{FORTRAN} evolutionary track interpolator \footnote{available at \url{http://www.astro.yale.edu/demarque/yystar.html} and last updated on May 2018} based on the Yale-Potsdam stellar isochrones \citep{spada13,spada17}, taking as inputs the effective temperature $T_{eff}$, metallicity $[Fe/H]$ and luminosity $L_*$. 
The uncertainties on stellar masses and radii are derived with a Monte Carlo approach, running again the interpolator in a 0.25 M$_\odot$ neighborhood centered on the M$_\odot$ obtained from the first run of the interpolator and using input values of T$_{eff}$, [Fe/H] and L$_\odot$ randomly drawn from a Gaussian distribution having standard distribution equal to each parameter's archive uncertainty. From this new run of the evolutionary track interpolator we therefore obtain distributions of M$_*$ and R$_*$, the standard deviations of which are then used as the 1-$\sigma$ uncertainties of stellar masses and radii.
For each star in the sample, we retrieve from the \emph{Gaia} DR2 archive values for $T_{eff}$ and parallax $\pi$; the latter is then used to compute stellar luminosity $L_\odot$.
Having no high-resolution near-infrared spectra available that would enable a reliable chemical characterization of the stars, we fix the input metallicities at $[Fe/H]=0$ dex; this choice is a reasonable assumption for the analysed stellar types and is further supported by a recent study on a similar sample of M dwarfs by \cite{newton2016} for which near-infrared spectra estimate an average iron abundance comparable to 0 dex.
Of the 247 stars in our sample, 27 are not found in the \emph{Gaia} DR2 archive and 3 have no \emph{Gaia} estimate of $T_{eff}$; for these stars the use of the Yale-Potsdam interpolator was therefore not possible.
In Fig.\,\ref{fig2} we show the mass distribution of our M dwarfs sample on which the period search was performed with the mass distribution of the subsample with measured rotation over-plotted (dash-dotted red line).
We perform the two-sided Kolmogorov-Smirnov (K-S) statistic test to evaluate the associated probability $p_\mathrm{K-S}$ that the two mass distributions are drawn from the same distribution function.
We obtain $p_\mathrm{K-S}=0.9999$, with therefore a high likelihood of the two mass distributions having been drawn from the same cumulative distribution function.
We conclude that there is not a clear bias in the period detection (see Sec.\,\ref{period_d}) with respect to the mass.

\begin{figure}
\includegraphics[width=1\linewidth]{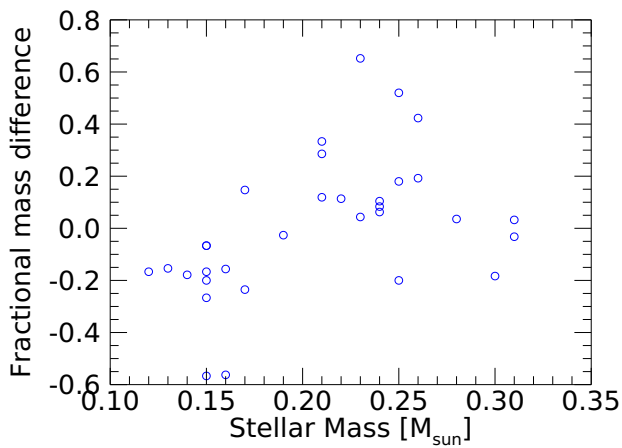}
\caption{Fractional mass difference between the masses derived by \protect\cite{irwin11} and the masses as derived in this work as a function of stellar mass. }
    \label{fig2bis}
\end{figure}

Since the uncertainty over the mass estimation ($\leq$ 10\%) heavily benefits from the \emph{Gaia} DR2 data, we re-derive, as described above, the values of mass and radius of the M dwarf sample with measured rotation from \cite{irwin11}.
We select this sample despite the fact that it is a sub-sample of more recent works like \cite{newton2016} and \cite{newton2018} because there are many similarities with the statistical properties of the APACHE sample, e.g. number of targets observed, number of photometric points per target, phase coverage, photometric single point uncertainties. Furthermore, the most recent works do not  drastically revise the statistics presented in the first paper, so we can use this sub-sample as a proxy of them.
In Fig.\,\ref{fig2bis} we present the fractional differences between the masses of the \cite{irwin11} sample as derived by the authors and the masses as derived in this work as function of stellar mass. It shows differences up to 60 \% and a slight systematic shift of our mass estimations towards bigger values for the less massive stars.


\section{Results}
\label{results}

\subsection{Period detection}
\label{period_d}

As described in Sec.\,\ref{diff_phot}, the output of our pipeline is a light curve where the vast majority of the systematic effects are properly accounted for. 
So, at the first order, we can assume that the variance of the final light curve is composed only by photon noise, correlated and uncorrelated stellar jitter and, if present, the stellar signals as the modulation induced by the presence of spots.
Therefore, we searched for sinusoidal-like modulation in the light curves without any other filters by using the complete dataset binned at 30 minutes. The bin width was selected in order to increase the SNR and to weaken the short time scale correlated noise while preserving a time sampling good enough to detect rotation periods $< 1$ days.  We used the GLS to calculate the frequency periodograms sampled on a uniform grid in frequency corresponding to the period interval between 0.1 and 500 days.
To estimate the significance of the detection, we performed a bootstrap analysis (with replacement) using 10000 randomly permuted data-sets derived from the original binned photometric data.
We select all the objects with FAP $\leq$ 1\%, while the final sample of periodic variables was done on the basis of a visual inspection.
One hundred eight light curves passed this selection, where the other 140 light curves were consistent with no detectable variation, had excessive residual systematics, or had insufficient phase coverage to determine the true period.

Here we note again that, in order to avoid the underestimation of the signal amplitude due to the de-trending algorithm, we correct for the systematics and fit the sinusoidal signal simultaneously, as described in Sec.\,\ref{diff_phot}, for all the targets in the final rotating sample.

\begin{figure*}
$\begin{array}{cc}
\includegraphics[width=0.54\linewidth]{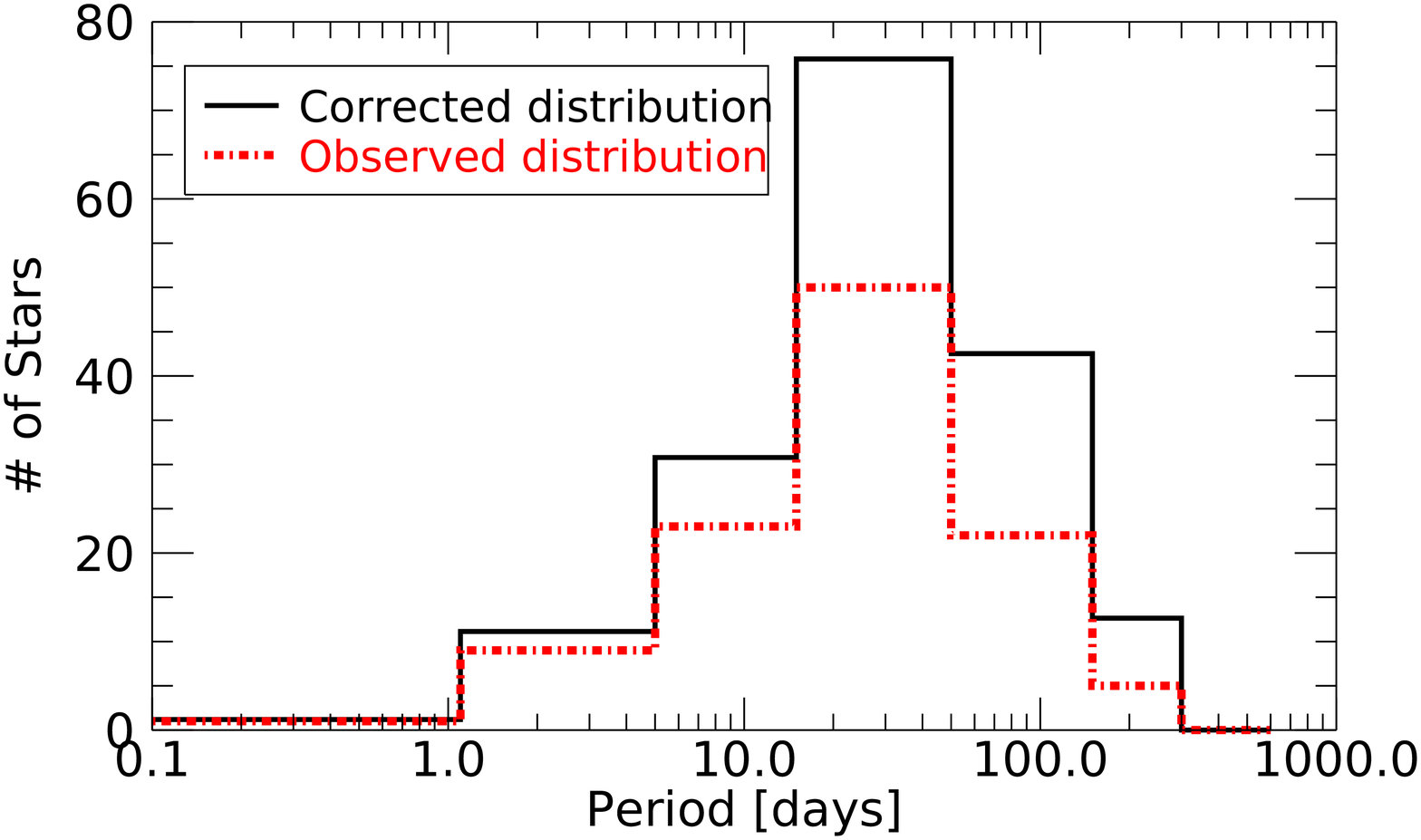} &
\includegraphics[width=0.44\linewidth]{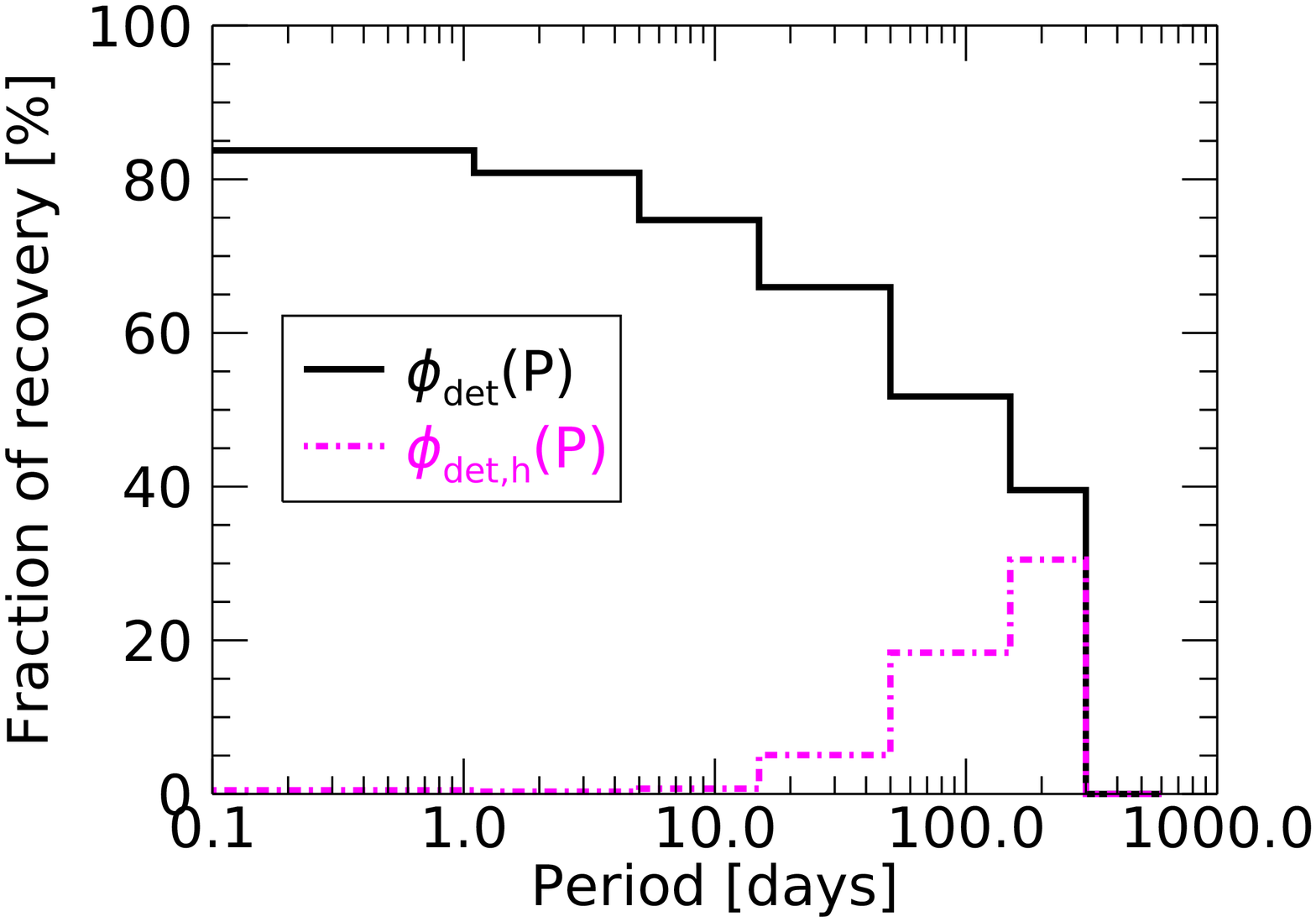} \\
\end{array} $
 \caption{Left panel: the solid black line represents the number of stars as a function of rotation period for the APACHE sample while the dash dotted red line represents the number of stars corrected for the detection efficiency as a function of rotation period. Right panel: the solid black line represents the detection efficiency $\phi_{det}(P)$ derived from the simulations as a function of rotation period while the dashed magenta line represents the harmonic contamination $\phi_{det,h}(P)$ derived from the simulation as a function of rotation period.}
\label{fig4}
\end{figure*}

\subsection{Rotation period distribution}
\label{period_distribution}
\begin{figure}
    \includegraphics[width=1\linewidth]{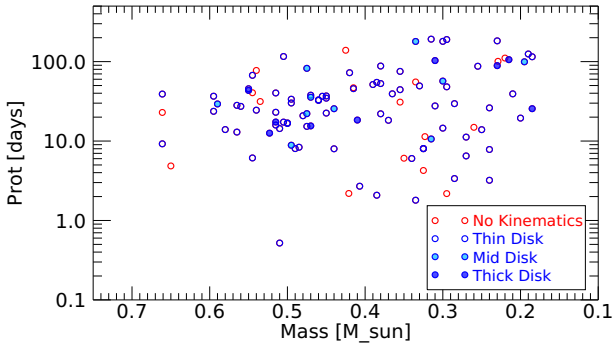}
    \caption{Rotation period as a function of stellar mass for the APACHE sample. The red circles represent the stars with no kinematics characterization. The blue circles represent the targets with kinematics characterization, where the unfilled circles are the stars more likely belonging to the thin disk, and the filled ones belonging to the thick disk. The filled light blue circles have an intermediate kinematics between the two populations.}
    \label{fig3bis}
\end{figure}

The rotation period distribution of the sample of 107 M dwarfs with detected rotational modulation is shown in the left panel of Fig.\,\ref{fig4}, while the results are presented in Table\,\ref{tab:par1}.
The distribution's peak lies at $\sim 30$ days with a minimum rotation period of $0.52$ days and a maximum of $191.82$ days.
Despite our sensitivity decreases at longer periods, in this range it is substantially unbiased, as shown in Sec.\,\ref{simul}.
Fig.\,\ref{fig3bis} shows rotation period plotted as a function of stellar mass, where the filled symbols indicate our kinematic population assignments from Sec.\,\ref{kinematic}.
Looking at Fig.\,\ref{fig3bis} and considering the upper periods envelope, the periods are observed to increase with decreasing mass.

\begin{figure}
    \includegraphics[width=1\linewidth]{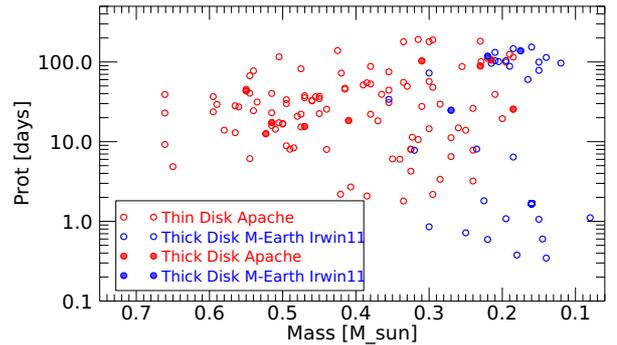}
    \caption{Rotation period as a function of stellar mass. The red circles represent the stars from the APACHE sample while the blue circles represent the stars from the MEarth \protect\cite{irwin11} sample. The filled circles are the stars more likely belonging to the thick disk. For clarity we do not show the stars with the intermediate kinematic but the two samples are in agreement.}
    \label{fig3tris}
\end{figure}

While the thin disk sample spans the full range of periods and includes entirely the rapidly rotating objects,
the thick disk sample is constituted by only slowly rotating stars.
The situation is similar to the thin disk sample for the stars with an intermediate kinematics classification, with no objects in the ultra fast ($P < 5$ days) rotating clump.
We discuss this point in the next Sections, but given the mean old age of the thick disk, this indicates that the older objects in the sample are rotating more slowly.
As shown in Fig.\,\ref{fig3tris}, our findings are in agreement with the \cite{irwin11} MEarth sample revised in mass and kinematics by this work.
In any case, the mean ages of the thin and thick disks only provide a weak metrics to understand the rotational evolution of the M dwarfs.

\subsection{Kinematics characterization (as a proxy of the mean age of the sample)}
\label{kinematic}
\begin{figure}
\includegraphics[width=1\linewidth]{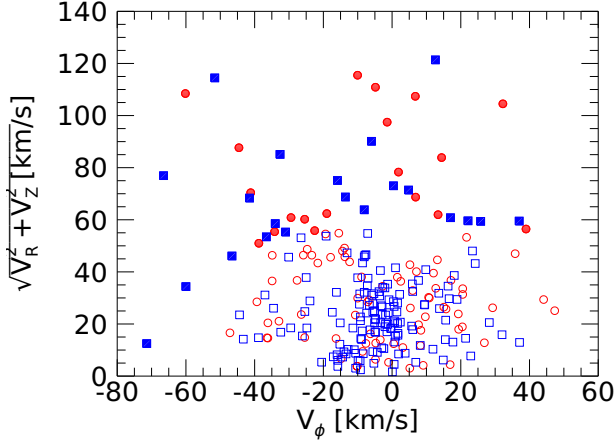}
\caption{$\sqrt{V^2_R + V^2_Z}$ as a function of $V_{\phi}$ for the APACHE sample with measured rotation (red circles) and the MEarth-N \protect\citep{newton2016} sample (blue square). The filled symbols are the stars associated with mid/thick disk.}
    \label{fig2tris}
\end{figure}

As discussed in the previous Section, the rotational period distribution is a function of mass and age. 
The age assignment for the field M dwarf stars is not trivial.
A tentative solution is to use the available kinematic information to infer a rough estimate of the stellar age.
For our sample, we consider the \emph{Gaia} DR2 astrometric parameters (positions, parallaxes and proper motions) and line-of-sight velocities, when available. When the line-of-sight velocity is not available from \emph{Gaia} DR2, we consider the reference value from SIMBAD\footnote{(http://simbad.u-strasbg.fr/simbad/)} and reference therein.
The distance to the stars is then calculated by naively inverting \emph{Gaia} DR2 parallaxes.
All the stars in our sample have $\varpi/\sigma_{\varpi} > 10$, indicating that the obtained distances are not prone to the biases associated with large fractional parallax error \citep{bailerjones15}.
We derive 3D positions and velocities in Galactocentric cylindrical coordinates $R,\phi, Z, V_R, V_{\phi}, V_Z$ where $R$ is the distance from the Galactic center, $\phi$ is the Galactic azimuth (defined as positive in the direction of Galactic rotation), $Z$  is the height from the Galactic plane and  $V_R, V_{\phi}, V_Z$ are the velocities components along the coordinates described above.
We assume for the Sun a distance to the Galactic center of $R_{\odot}=8.122$ kpc \citep{gravitycol18}, a vertical height above the Galactic midplane $Z_{\odot}=0.027$ kpc \citep{chen01} and  a Galactic azimuth of the Sun $\phi_{\odot} = 0$.
We assume for the Sun a velocity $(V_{R,\odot},V_{\phi,\odot},V_{Z,\odot}) = (-12.9,245.6,7.78) $ km/s  \citep{drimmel18}.
Kinematic information can be used to estimate the mean stellar age of the sample by statistically assigning our targets to the Galactic thin disk or thick disk/halo populations.
In particular, the Galactic thick disk is characterized by high-velocity dispersions, a unique chemistry and remarkably old age.
For the stars of our sample, we therefore calculate the total velocity with respect to the Local Standard of Rest (LSR) $V_{TOT}$ as
\begin{equation}
\label{vtot}
    V_{TOT}= \sqrt{V^2_R + (V_{\phi} - V_{LSR})^2+ V^2_Z}
\end{equation}
with $V_{LSR}=233.4$ km/s \citep{drimmel18,schonrich10}.
We identify thick disk star candidates as the ones with $80$ km/s $< V_{TOT} < 180$ km/s (following \citealt{nissen04}), whereas thin disk stars are expected to have $V_{TOT} < 60$ km/s \citep{nissen04}. We define a mid disk population as all those stars with the intermediate $60$ km/s $< V_{TOT} < 80$ km/s kinematic. 
According to this method, our sample with measured rotation contains $\sim$ 80\% of thin disk stars, $\sim$ 10\% of mid disk stars, while $\sim$ 10\% of stars are thick disk stars. 
This is in agreement with the stellar number density distribution of the Milky Way from the Sloan Digital Sky Survey (SDSS) in the Solar neighborhood \citep{juric08}.
In Fig.\,\ref{fig2tris} we show the Toomre diagram (e.g. \citealt{gaiacol18}) of the APACHE sample in red and the \cite{newton2016} sample in blue, using the velocity vector components published in \cite{newton2016} for the MEarth sample. The filled symbols are the stars associated with mid/thick disk. 
In \cite{newton2016} they found that 14\% $\pm$ 3\%  of ``grade A'' rotators likely belonged to the thick disk, a number becoming 7\% $\pm$ 3\% considering the grade A + B rotators. No stars with detected rotation were found to belong to the halo populations. 
On this basis, we can consider the two samples as equivalent in terms of kinematic populations, and, consequently, equivalent in terms of age distribution, according to the kinematic population/age relation presented in \cite{newton2016} and discussed here. This is not an unexpected result, considering that the samples are selected in the solar neighborhood.

Finally, we assign to the thin disk stars a mean age of $\sim 3$ Gyr while to the thick disk stars a mean age of $\sim 10$ Gyr (e.g., \cite{feltzing08}).

\subsubsection{The age-velocity dispersion relationship} 

As well observed in the open clusters at different ages (e.g., \citealt{herbst02},  \citealt{hartman09b}, \citealt{prosser95},), the low-mass main-sequence stars spin down with time.
Therefore, as highlighted in Sec.\ref{period_distribution}, it is expected that slow rotators (the upper envelope in the rotation period-mass relation) are older than their more rapidly rotating counterparts. 
While in clusters there are robust method to constrain the ages and therefore the rotational evolution of low mass stars over a grid of young ages, there are no reliable methods to determine the ages of isolated field M dwarfs.

\begin{figure}
    \includegraphics[width=1\linewidth]{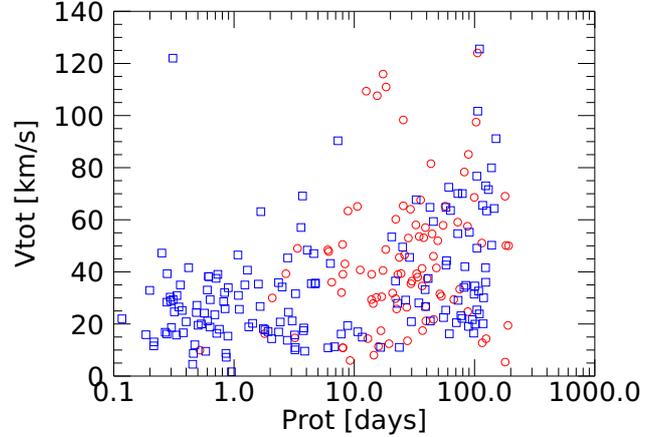}
    \caption{Total space velocity $V_{tot}$ as a function of rotation periods.The red circles represent the stars from the APACHE sample while the blue squares represent the stars from the MEarth \protect\cite{newton2016} sample.}
    \label{fig3trisbis}
\end{figure}

Looking at the age-velocity dispersion relation in the solar neighbourhood (e.g. \citealt{yu18}, \citealt{aumer09}), we investigate the signature of an age-rotation relation in the distribution of total space velocities, as defined in Eq.\ref{vtot}, as a function of photometric rotation period.
Fig.\,\ref{fig3trisbis} shows the total space velocity as a function of measured photometric rotation period, where the blue circles are the stars from the APACHE sample while the blue circles represent the stars from the MEarth-N \protect\cite{newton2016} sample.
We use the Spearman rank correlation coefficient for total space velocity $V_{TOT}$ and rotation period which is $0.22$ for APACHE sample while the Spearman rank correlation coefficient published by \cite{newton2016} for the full sample of MEarth late M dwarfs is $0.18$, a value in good agreement with that derived for the APACHE sample.
Assuming that the velocity dispersion increases with age, as expected for an older stellar population that has been dynamically heated, the star's ages are increasing with rotation period.

In order to adopt the age-velocity relation as published in \cite{yu18} or \cite{aumer09}, we need to estimate the dispersion of the $V_Z$ velocity component, $\sigma_{V_{Z}}$.
We determine the $\sigma_{V_{Z}}$ that underlies our data closely following the Bayesian approach described in \cite{newton2016}. 
Following their discussion, we divide our sample into bins in period, $P < 1$ day, $1 < P < 10$ days, $10 < P < 70$ days, and $P > 70$ days.
In Tab.\,\ref{tab3} we summarize our findings, compared to the results by \cite{newton2016}. We use the age-velocity relation as published in \cite{aumer09} in order to compare our results with the MEarth survey. In any case, the considerations presented here, based on a relatively small sample and a large intrinsic error, are not affected by the used relation.
It is useful here to remember that the APACHE and MEarth samples do not cover the same mass range but rather can be considered as complementary across the M dwarf mass range. We exploit this in order to track the M dwarfs' rotational evolution path and relative time scales as a function of the stellar mass.

\begin{table}
	\centering
	\caption{Velocity dispersion $\sigma_{V_Z}$ and estimated ages for M dwarfs with detected rotation periods from APACHE and MEarth-N. The values reported here for the MEarth surveys are published in \protect\cite{newton2016}, Table 6}
	\label{tab3}
	\begin{tabular}{lccccr} 
		\hline
		Period Bin & N stars & Mean P & $\sigma_{V_Z}$ & Est. Age \\
		 (Days) &  \# & (days) & (km/s)  & Gyr\\  
		\hline
		APACHE &  &  & & \\
		\hline
        $0.1 < P < 1$ & 1  & 0.52   & ... & ... \\
        $ 1 < P < 10$ & 16 & 6.14   & $19.9^{+3.8}_{-10.2}$    & ... \\
        $10 < P < 70$ & 58 & 29.63  & $14.6^{+2.3}_{-2.3}$     & $3.3^{+0.6}_{-0.6}$ \\
        $P > 70$      & 17 & 122.49 & $20.1^{+3.4}_{-10.5}$    & ... \\
		
		MEarth-N &  &  &  &  \\
		\hline
		$0.1 < P < 1$ & 39 & 0.5   & $6.0^{+1.8}_{-1.0}$ & $0.5^{+0.4}_{-0.2}$\\
        $ 1 < P < 10$ & 23 & 2.9   & $7.4^{+1.8}_{-1.8}$ & $0.7^{+0.5}_{-0.3}$\\
        $10 < P < 70$ & 10 & 28.3  & $6.5^{+1.6}_{-1.5}$ & ...\\
        $P > 70$      & 14 & 102.4 & $16.7^{+5.3}_{-4.5}$ & $4.5^{+3.9}_{-2.3}$\\
	\end{tabular}
\end{table}

For the APACHE sample the only period's bin well sampled is in the range $ 10 < P < 70$. For it, we adopt a mean age of $3.3 \pm 0.6 $ Gyr. 
In this bin, our mean age is comparable with the mean age in the $P > 70 $ range for the MEarth sample. This is a further proof of how the upper envelope of the period-mass relation (see Sec.\,\ref{kepler} and Sec.\,\ref{compare_me} for further details) is a function of mass with the late mature M dwarfs rotating slower in comparison with the early mature M dwarfs. 
In the other bins our results are not robust enough to assign a mean age.

\subsection{Amplitude of variability}

\begin{figure}
    \includegraphics[width=1\linewidth]{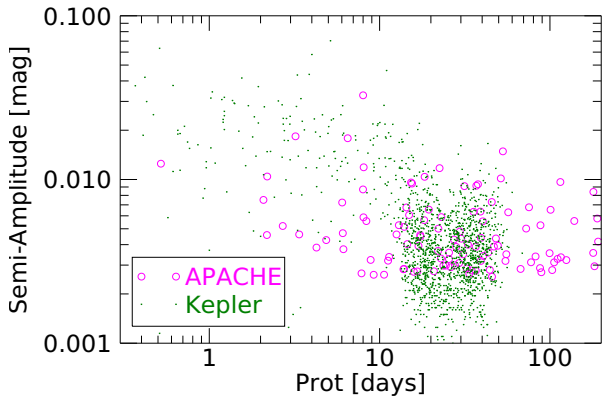}
    \caption{Semi-amplitude of the signal as a function of rotation period. The red circles represent the stars from the APACHE sample while the green dots represent the stars from the \textit{Kepler} \protect\cite{mcquillan13} sample.}
    \label{fig3quadris}
\end{figure}

The amplitude of the periodic photometric modulation associated to the stellar rotation depends to the contrast between the spotted and unspotted stellar photosphere and the longitudinal inhomogeneity in the distribution of spots.
Since the starspots are related with the stellar activity, which is in turn related to stellar age, the stellar rotation and the variation of the magnetic fields, we investigate the correlation between rotation period, mass and photometric amplitude of the signals.
Fig.\,\ref{fig3quadris} shows the semi-amplitude of the signal as a function of rotation period. There is not a clear correlation between the amplitude of the signals and the rotation period. This is confirmed by \cite{mcquillan13} for M dwarfs in the Kepler sample (we discuss more in details the comparison between APACHE and \textit{Kepler} in Sec.\,\ref{kepler}) although  they observed a bigger dispersion of the amplitude in the fast rotators sample.
It is important to note that the amplitude of the periodic signals of the Kepler stars could be reduced by the MAP pipeline used to reduce the data analysed by \cite{mcquillan13}, in particular for periods longer than 30-40 days (see Fig. 6 in \cite{gilliland15}).


In any case, considering that the amplitude of the flux modulation depends also on the inclination of the star spin axis, we do not exclude that the amplitudes distribution may be dominated by a sort of geometrical effect.
Furthermore, we do not model any evolution or non-sinusoidal behavior. In these cases the ``global'' semi-amplitude is suppressed relative to the peak-to-peak amplitude that was instead measured by \cite{mcquillan13}.

\subsection{Simulation}
\label{simul}

In order to evaluate our sensitivity in period and amplitude, simulations were performed using the following method: for each target we inject 10000 sinusoids with periods from 0.1 to 200 days following a uniform distribution in frequencies.
For each period, we randomly extract the reference phase in the range between 0 and 1.
A fixed semi-amplitude was adopted in our simulations, 0.002 mag, corresponding to the minimum amplitude of our rotation candidates (see Fig.\,\ref{fig6} or Tab.\,\ref{tab:par1}).
We inject the synthetic signals into the 140 light curves with no period detection, as described in the previous section, while we made synthetic light curves for the other targets. 
For each synthetic light curve we randomly add noise generated according to a Gaussian distribution with sigma equal to the mean single point uncertainty derived from the real data.
Each simulated light curve was processed exactly in the same way of the real data, as described in Sec.\,\ref{period_d}.
We have a detection if the retrieved period differs from the true injected period by $\leq$ 1\%.
We define the detection efficiency $\phi_{det}(P)$ as the relative number of periods that are detected by GLS with respect to the total number of injected periods.
Although we did not consider as a detection the multiple and the sub-multiple periods of the true injected period, we can use them to investigate the possible contamination by the harmonics in our period distribution. 
We have a detection of an harmonic or sub-harmonic (up to the fourth harmonic or subharmonic) if the retrieved period differs from the corresponding harmonic or sub-harmonic of the injected period by less than 1\%.
We define the harmonic contamination $\phi_{det,h}(P)$ as the relative number of harmonics that are detected by GLS with respect to the total number of injected periods.

Our estimation of detection efficiencies allows us to investigate the presence or not of bias into our period distributions due to the single point uncertainty, the time sampling and the phase coverage of the data.
We correct the measured period distribution with $\phi_{det}(P)$: due to the phase coverage no clear bias is evident. The corrected period distribution is shown in the right panel of Fig.\,\ref{fig4}. 
We highlight that this  ``corrected'' distribution does not account for bias in masses or activity levels but only for the survey sensitivity and, summarizing, it is useful to quantitatively address the quality of our data and the position of the distribution peak under the hypothesis of the initial sample selection.

The result in terms of recovery rate, as well the harmonics' contamination, is summarized in the left panel of Fig.\,\ref{fig4}.
The statistics indicate very good period recovery, 80\%-60\%, for the bins between 0.1-100 days  while it drops in longest-period bin, where the completeness reduces to $\sim$ 40\%. On the other side, the harmonics' contamination is 0\% for the bins between 0.1-15 days  while it grows up in the longest-period bin, where the contamination increases to $\sim$ 30\%.
This is expected due to the limited survey duration and the intrinsic difficulties to take into account annual trends due to the instrumental instability. 
It is interesting to note that the 98\% of the harmonics' contamination is due to the half or the double of the true injected period.

For each target, we evaluate a mean detection efficiency $\Phi_{det}$ and a mean harmonics' contamination $\Phi_{det,h}$ in the whole range 0.1 to 200 days and we report these  values in Tab.\,\ref{tab:par2}.

$\Phi_{det}$ and $\Phi_{det,h}$ could be useful, in certain cases, to try to break down the degeneracy between the true period and its harmonics (see Sec.\,\ref{spec} for some examples)  due to the phase coverage and time sampling. Furthermore, it could be considered as a further check on the reliability of the periods published in this work.

\section{The APACHE period distribution in a wide context}
\label{secwide}
\subsection{Spectroscopic rotation periods}
\label{spec}

As part of a joint exoplanet-hunt effort, as discussed in Sec.\,\ref{aic}, 44 of the APACHE targets were also monitored spectroscopically by the HADES program. The HADES targets were investigated for the presence of rotation signals in the spectroscopic data by \cite{suarezmascarenoetal2018}, who studied the Ca~{\sc ii}  H and K and H$\alpha$ chromospheric activity indicators, the RV series, and the parameters of the cross correlation function.
We detected significant periodic variability in the APACHE photometry for 23 of these 44 stars, while the other 21 lightcurves did not pass the quality checks described in Sec. \ref{period_d}.
For 13 stars over these 23, \cite{suarezmascarenoetal2018} were able to measure the rotation periods from the spectroscopic activity indexes (Ca II H\&K and/or H$\alpha$) or RV time series, while for the other 10 the rotation period was estimated from activity-rotation relationships and their calculated values of $\log R'_\text{HK}$. The definition of $\log R'_\text{HK}$ was extended for application on M dwarfs spectra, following a procedure very similar to the one used by \cite{suarezmascarenoetal2016}.
We summarize the properties of this subsamble in Tab.\,\ref{tab_hade}.

\begin{table}
	\centering
	\caption{Rotational periods of the 23 Apache targets in common with \protect\cite{suarezmascarenoetal2018}. In the $P_{rot,H}$ sources column we denote the methods to derive the rotation periods from the spectroscopic data.}
	\label{tab_hade}
	\begin{tabular}{lccccr} 
		\hline
		Name       &      $P_{rot,A}$     & $P_{rot,H}$      &  $P_{rot,H}$ sources \\

           &     (days)      & (days)     &                  \\
          
		\hline

J06147+4727     &     27.320  &  $ 23.0 \pm  4.0 $  &    estimated     \\   
J07316+6201W    &     67.240  &  $ 15.0 \pm  3.0 $  &    estimated     \\
J09133+6852     &     19.450  &  $ 10.4 \pm  0.1 $  &    Act ind	 \\
J01013+6121     &     75.350  &  $ 34.7 \pm  0.1 $  &    Act ind, RV   \\
J12350+0949     &     18.260  &  $ 55.0 \pm  5.5 $  &    Act ind	 \\
J17158+1900     &     14.540  &  $ 37.0 \pm 13.0 $  &    Act ind	 \\
J17166+0803     &     45.690  &  $ 85.0 \pm 15.0 $  &    estimated     \\
J18596+0759     &     43.400  &  $ 25.0 \pm  5.0 $  &    estimated     \\
J21129+3107E    &     20.830  &  $ 18.0 \pm  4.0 $  &    estimated     \\
J21185+3014     &      8.060  &  $ 7.80 \pm  0.2 $  &    Act ind, RV   \\
J12194+2822     &     22.910  &  $ 23.2 \pm  0.1 $  &    Act ind, RV   \\
J16254+5418     &     87.969  &  $ 77.8 \pm  5.5 $  &    Act ind, RV   \\
J14294+1531     &     12.570  &  $ 43.5 \pm  1.5 $  &    Act ind	 \\
J18353+4544     &     33.620  &  $ 34.5 \pm  4.7 $  &    Act ind	 \\
J17160+1103     &     32.990  &  $ 33.6 \pm  3.6 $  &    Act ind, RV   \\
J11000+2249     &    179.920  &  $ 58.0 \pm 10.  $  &    estimated     \\
J11511+3516     &     36.980  &  $ 40.0 \pm  8.0 $  &    estimated     \\
J14257+2337W    &     17.360  &  $ 36.6 \pm  0.1 $  &    Act ind	 \\
GJ3649          &     30.150  &  $ 15.0 \pm  3.0 $  &    estimated     \\
J02565+5526S    &    103.100  &  $ 51.2 \pm  4.4 $  &    Act ind       \\
J03437+1640     &     22.490  &  $ 25.0 \pm  5.0 $  &    estimated     \\
J04086+3338     &      8.360  &  $ 32.4 \pm  1.6 $  &    Act ind, RV   \\
J04587+5056     &      7.990  &  $ 25.0 \pm  5.0 $  &    estimated     \\
        
	\end{tabular}
\end{table}

\begin{figure}
   \centering
 \includegraphics[width=1\linewidth]{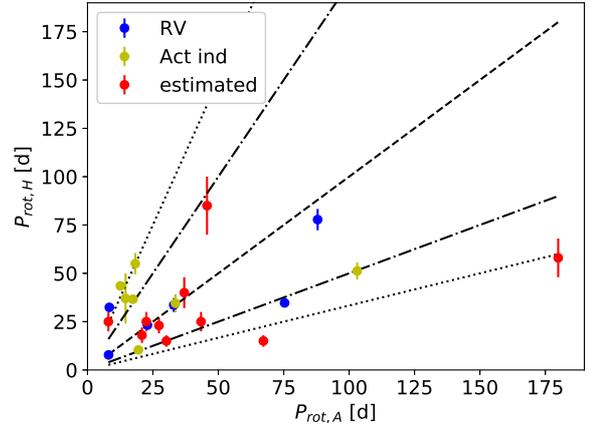}
 \caption{Comparison between the photometrically-derived rotation periods from APACHE $P_{rot,A}$ in days and the spectroscopically-derived ones from HADES $P_{rot,H}$ in days. The blue dots represent the targets with measured HADES rotation periods from RV time series, the yellow dots represent the targets with spectroscopic rotation periods derived from the activity index time series while the red dots represent the targets with spectroscopic rotation periods estimated from activity-rotation relationships. The dashed black line represents the $P_\text{rot, H} = P_\text{rot, A}$, the dash-dotted lines represent where one rotation period is either double or one half of the other while the dotted lines represent where one rotation period is either triple or one third of the other.}
 \label{fig5}
\end{figure}

Fig.\,\ref{fig5} shows the comparison between the APACHE and HADES rotation periods.
We can see that for most stars there is a good correspondence between either the two rotation periods or one rotation period and the first harmonic of the other. It is worth noticing that only for five targets there is a larger discrepancy:  two estimated HADES rotation periods on the lower-right corner of the plot appear to be largely underestimated with respect to the measured photometric periods, with one close to the second harmonic of $P_\text{rot, A}$; on the upper-left side of the plot, instead, there are three HADES measured periods which appear to be overestimated, and close to three times the photometric period. There is no apparent reason for these discrepancy with respect of the rest of the sample, as these stars do not appear to be either particularly active or quiet, and follow the general distribution of stellar parameters.

However, the photometric stellar signal at the rotation period is caused by stellar spots.
It is therefore important to note the existence of a ``pathological'' spot distribution (e.g two identical active regions longitudes spaced by $180^\circ$) which can lead to doubling of the frequency, causing to misestimate the true rotation period of the star \citep{ccameron09}.
Furthermore, the period estimations from activity-rotation relationships are affected by an uncertainty by a factor of 2-3 at a given level of activity (e.g, \cite{suarezmascareno15}; \cite{astudillo17}). 
The level of activity, in fact, changes due to activity cycles and the evolution of active regions. One should follow a star for a time comparable with its activity cycle to be sure that the average level of activity is correctly determined.
For this reason and from the nature of the activity-rotation relationships, which is precisely an empirical relationship and not a direct measurement, we suggest that these disagreements are not a cause of concern,  despite all the other uncertainties described in this Section.

Looking at the particular cases, the periods from activity-rotation relationships that are not on the 1:1 line do not seem to be randomly distributed in the diagram, as expected in the case of a wrong estimation.
There are two cases, J17166+0803 and J04587+5056, where we found double or triple periods of the photometric period.
Looking at the simulations, there is a low probability to have an harmonic contamination for these two cases.
Taking into account that the probability to have an harmonic contamination is drastically lower than the probability to have a sub-harmonic contamination, a similar conclusions could be done for the rest of the targets that do not lie on the 1:1 line.
We highlight here that our detection efficiency and harmonic contamination are calculated under the assumption that the signal induced by the rotation could be modelled with a sinusoid. We are not able to asses here the impact of this assumption on the detection efficiency.

Double or triple periods of the photometric period are found for different stars in the case of measurements based on chromospheric indices.
For these objects, we have a low probability harmonics contamination from our simulation. Taking into account that they are quite long periods, between 30 and 50 days, they could be associated with the timescale of evolution of the active regions that could dominate over the rotational modulation signal.

If we especially consider the periods obtained with the RVs, they are in agreement with the photometric ones in four cases out of six.
In one case, J01013+6121, the period from the RV is close to the first harmonic and this is observed in certain cases, even in the Sun as a star \citep{mortier17}. Looking at a $\Phi_{det} = 61\%$ and a $\Phi_{det,H} = 0\%$ (see Sec.\,\ref{simul} for the details), there are no reasons to suppose that the APACHE period estimation for this target is an harmonic of the true period. 
The only anomalous case, J04086+3338, is when the RV measurement is close to four times the photometric period.
In particular, we estimate a rotation period of 8.36 days with a semi amplitude of  0.0056 mag, while the measured HADES rotation period is 32.4 $\pm$ 1.6 days. 
From our simulation, we have a $ \Phi_{det} = 21\%$  and a $\Phi_{det,H} = 17\%$, that denotes a quite low detection efficiency for periods longer than 15 days and a consequent contamination of the true period by the harmonics.
In this case, we probably miss the true period due to the low temporal sampling in the APACHE photometry.
Generally speaking, such a type of discrepancies could be also explained as an effect of the "noise" necessarily present in the RV series, since the star is quite active and rotates in less than 10 days (according to photometry).

Focusing on the HADES targets with detected planetary systems which were present in the final sample of APACHE rotation periods, GJ 3998 \citep{afferetal2016} and GJ 625 \citep{suarezmascarenoetal2017}, we see a good agreement between the published spectroscopic rotation periods and the photometric values presented in this work. Moreover, we notice that for GJ 685 we do not present here any rotation period since the stellar rotation signal is very weak in the APACHE photometry \citep{pinamontietal2019}, and it did not pass the quality checks of the present analysis.

Our comparison with the rotation periods derived by the spectroscopy suggests that a contamination between the true period and its harmonics could be possible. The only way to break down this degeneracy is to acquire more data, both spectroscopic and photometric. In any case, although the true periods estimation is indeed what one aims for, detection of the harmonics of the true period is still a useful piece of information. For the vast majority of the objects presented here the rotation period estimate based on APACHE photometry is the only measurement available. These data therefore could represent an important reference point for further investigations, such as those aiming at spectroscopic follow-up of small-radius transiting planets below the APACHE sensitivity but that might be detected by space-based photometric programs (e.g., TESS, CHEOPS).

\subsection{Comparison to \textit{Kepler} photometry}
\label{kepler}
The large sample of rotating M dwarfs from the \textit{Kepler} survey published in \cite{mcquillan13} is ideal to compare with the APACHE sample. \cite{mcquillan13} measured the rotation periods of 1570 M dwarfs (of the 2483 stars examined) in the mass range $0.3-0.55$ $M_\odot$, corresponding to the mass range of APACHE. 
As shown in Fig.\,\ref{fig8}, the periods distributions from \textit{Kepler} and APACHE are in good agreement.
We perform another K-S test to evaluate the associated probability that the two distributions are drawn from the same distribution function. 
We obtain $p_\mathrm{K-S}=0.883$, indicating that \textit{Kepler} and APACHE distributions are consistent with the samples being drawn from the same cumulative distribution function.
As shown in Fig.\,\ref{fig8}, if we consider a denser binning in order to reproduce the Fig 9 of \cite{mcquillan13}, we note that the two distributions are different aside two details:
i) while the \textit{Kepler} period distribution is clearly bimodal, with peaks at $\sim$ 19 and $\sim$ 33 day, the APACHE period distribution does not show it clearly. \cite{mcquillan13} supposed that this bimodal behaviour hints at two distinct waves of star formation, a hypothesis that could not be true for our sample. Furthermore, as shown in Sec.\,\ref{spec}, we often observe degeneracy between multiples, sub-multiples and the true periods. Taking into account the smaller dimension of our sample combined with this effect, we are probably not able to resolve this bimodality, if present;
ii) \cite{mcquillan13} did not detect rotation periods longer than 70 days in any of their objects, although they searched as long as 155 days. Probably, this is mostly due to the initial selection of the Kepler sample which virtually does not include fully convective stars and therefore should not have periods longer than $~$70 days according to the observed mass-period relation.
On the other hand, we can not exclude that this discrepancy could be also explained as a combination of at least three effects: 1) it is possible that Kepler's systematics, particularly due to the rotation of the satellite every $\sim$ 90 days, affect the recovery of longer rotation periods. Furthermore, \cite{gilliland15} shows that the sensitivity for periods $\geq$ 15-20 days is lower than expected; 2) as shown in Sec.\,\ref{spec}, the degeneracy between multiples and submultiples of the true period could also affect the statistic on the longer periods; 3) \cite{mcquillan13} estimated masses from the Kepler Input Catalogue (KIC) effective temperatures without precise parallaxes. Since parallaxes and $T_{eff}$ are not available in DR2 for the majority of the \textit{Kepler} M dwarfs, it is not possible to derive the masses with the same method described in \ref{sample}.
Therefore, there may be an offset between the two mass scales, as also shown in \cite{berger18}.
Since the rotation periods become longer for older stars, as shown in \cite{newton2016} and in \cite{mcquillan13}, where the slope of the upper envelope of the period-mass relation changes sign around 0.55 $M_\odot$, below which period rises with decreasing mass, an uncertainty (or a systematic shift) of $\sim$20\%  on mass could match these disagreement.

\begin{figure}
    \includegraphics[width=1\linewidth]{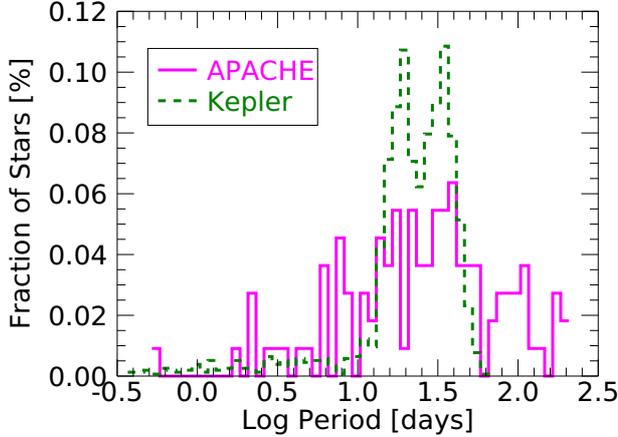}
    \caption{Fraction of stars as a function of the logarithm of the rotation period (expressed in days). The red solid line draw the APACHE period distribution while the green dashed line draw the Kepler period distribution.}
    \label{fig6bis}
\end{figure}
 
\begin{figure}
    \includegraphics[width=1\linewidth]{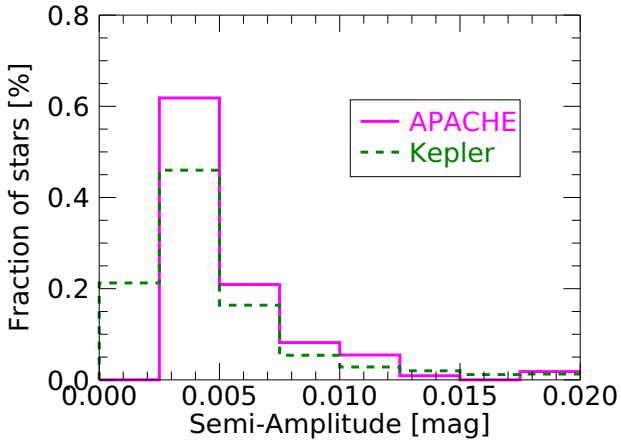}
    \caption{Fraction of stars as a function of signal semi-amplitude in magnitude. The red solid line drawn the APACHE distribution while the green dashed line drawn the Kepler distribution.}
    \label{fig6}
\end{figure}

In Fig.\,\ref{fig6}, we draw the signal amplitudes distributions from \textit{Kepler} and APACHE rotators. 
Looking at the APACHE distribution, we note how we do not observe semi-amplitudes below 2.5 mmag. This is probably due to the survey sensitivity limit over an observation season.
Based on a new K-S test, we obtain $p_\mathrm{K-S}=0.815$, showing that \textit{Kepler} and APACHE semi-amplitude distributions are consistent with the samples being drawn from the same cumulative distribution function.
It is important to note that the photometric band-pass of the \textit{Kepler} survey is different from that of the APACHE survey. The direct consequence of observing at different wavelengths is to obtain different contrast between spotted and unspotted photosphere and, consequently, a different amplitude of the modulation. 

\subsection{Comparison with rotation periods from MEarth North \& South}
\label{compare_me}
Since in Sec.\,\ref{kinematic}, we used the MEarth \cite{irwin11} sample to show how the properties in terms of kinematic of the two samples are substantially equivalent, here we use the complete sample of late type M dwarfs from \cite{newton2016} and \cite{newton2018} to compare the late M dwarfs mass-period relation from MEarth to the early M dwarfs mass period relation from APACHE and Kepler surveys.
Since APACHE and  Kepler surveys both have a star mass distribution that peaks at 0.5 $M_\odot$, while MEarth has a mass distribution that peaks at 0.2 $M_\odot$, this is a good opportunity to bridge the two samples over the full range of M dwarf masses.
As highlighted in Sec.\,\ref{mas}, the masses for MEarth and Kepler are determined using different methods with respect to this work and do not benefit of the \emph{Gaia} DR2 data, so there may be an offset between the two mass scales and the APACHE one. In any case, an exact review of the masses is beyond the scope of this work and these uncertainties do not heavily affect the statistical considerations presented here.

In Tab.\,\ref{tab2} we summarize the property of APACHE, \textit{Kepler} and MEarth samples. 
\begin{table*}
	\centering
	\caption{Summary of rotation sample properties from APACHE, \textit{Kepler} and MEarth surveys}
	\label{tab2}
	\begin{tabular}{lccccr} 
		\hline
		Survey & Reference paper & Stars with measured rotation & Stars sample & Period Range & Mass Range\\
		 & & \# & \# & (days) & ($\mathrm{M_\odot}$)\\  
		\hline
		APACHE & this work  & 107 & 247 & 0.1-135 & 0.18-0.79\\
		\textit{Kepler} & \cite{mcquillan13} & 1570 & 2483 & 0.37-67 & 0.3-0.55\\
		MEarth-N & \cite{irwin11}    & 41  &  273  &  0.28-154 & 0.025-0.35 \\
		MEarth-N & \cite{newton2016} & 387 &  1883 &  0.1-140  & 0.06-0.78 \\
		MEarth-S & \cite{newton2018} & 234 &  574  &  0.1-149  & 0.08-0.38 \\
		\hline
	\end{tabular}
\end{table*}

\begin{figure*}
    \includegraphics[width=1\linewidth]{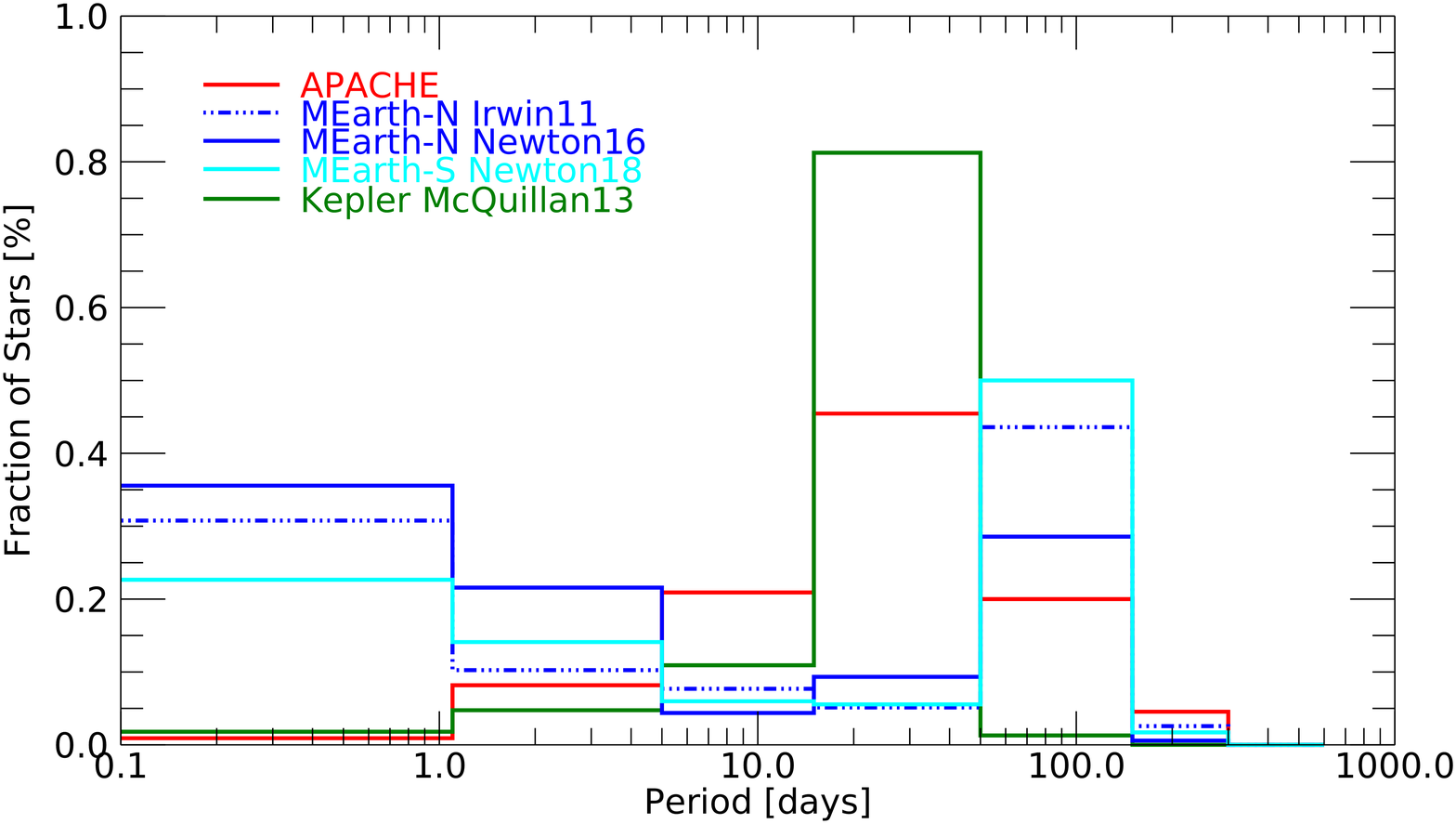}
    \caption{Fraction of stars as a function of rotation period. The red solid line draws the APACHE distribution, the blue dash-dotted line indicates the \protect\cite{irwin11} period distribution, the blue solid line draws the MEarth-North period distribution from \protect\cite{newton2016}, the light blue solid line indicates the MEarth-South period distribution from \protect\cite{newton2018} while the green solid line draws the \textit{Kepler} period distribution from \protect\cite{mcquillan13}.}
    \label{fig8}
\end{figure*}

In Fig.\,\ref{fig8} we show the rotation period distribution for the APACHE, \textit{Kepler} and MEarth surveys.
While the APACHE sample, supported by \textit{Kepler} data, shows a single peak around $\sim 30$ days, the MEarth period distribution appears to be in two clumps, a population of rapidly rotating objects with periods of $\approx$ 0.2-10 days, and a population of slowly rotating objects with periods of $\approx$ 30-160 days.
The APACHE period distribution has a bigger variance with respect to the \textit{Kepler} or MEarth slow rotation period distributions, due to the bigger variance of the APACHE mass distribution.
It is interesting to note in Fig.\,\ref{fig8} that while for the early M dwarfs samples from APACHE and \textit{Kepler} the shorter periods look like a tail in the period distributions, we see a gap between ``slow'' and ``fast'' rotators in the period distributions for the late M dwarf stars from MEarth. 

\begin{figure*}
    \includegraphics[width=1\linewidth]{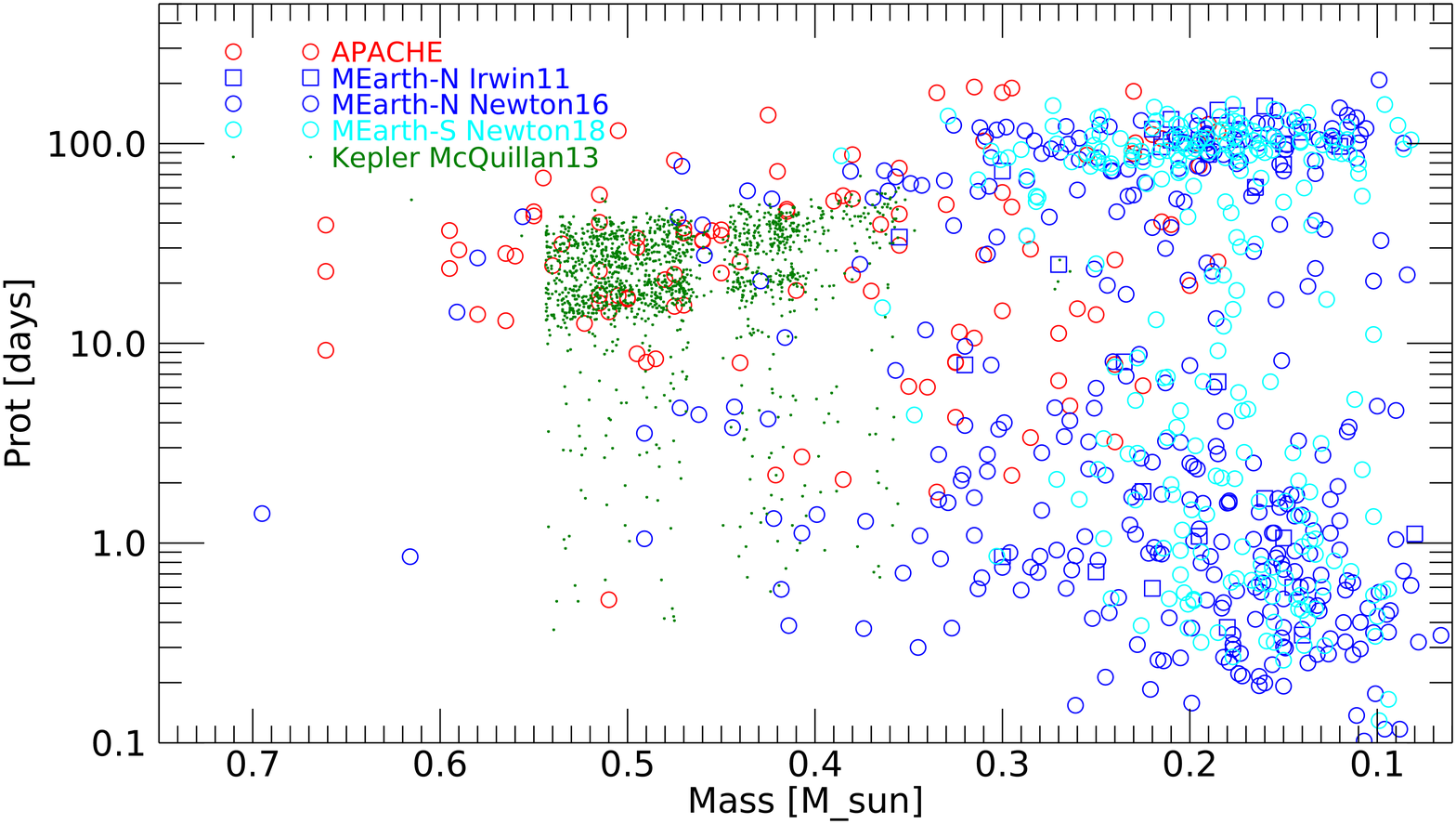}
    \caption{Rotation period as a function of stellar mass. The red circles represent the APACHE data, the blue squares represent the \protect\cite{irwin11} data, the blue circles the MEarth-North data from \protect\cite{newton2016}, the light blue circles represent the MEarth-South period distribution from \protect\cite{newton2018} while the green dots represent the \textit{Kepler} data from \protect\cite{mcquillan13}.}
    \label{fig7}
\end{figure*}

In Fig.\,\ref{fig7} we show the rotation period as a function of the stellar masses for the APACHE, \textit{Kepler} and MEarth surveys. First of all, there is a clear dependence on the stellar mass for slow rotators over the full mass range, with the lowest-mass stars reaching the longest rotation periods.
Furthermore, the two clumps in the MEarth periods distribution described above are here well visible in the mass range 0.1-0.3 $M_\odot$ where $\sim$ 50\% of the lowest-mass stars show the fastest rotation rates. Moreover, the gap between ``slow'' and ``fast'' rotators increases with decreasing mass.

The rotation periods we find for the APACHE M dwarfs are consistent with the MEarth results for the older/slow rotators stars and in the overlapping mass range between APACHE and MEarth. The lack of fast rotators between APACHE/\textit{Kepler} and MEarth could be principally due to the fact that APACHE and \textit{Kepler} do not observe M dwarf stars below the full convection limit around $\sim$ 0.3 $M_\odot$.

As described in Sec.\,\ref{aic}, during the initial target selection we cross-correlated our sample with many catalogs, searching for additional information such as measurements of projected rotational velocity \textit{V$\sin$i} and level of chromospheric activity and X-ray emission, to flag active targets.
We suggest that the impact of this initial selection on the observational strategy was very limited for two reasons: 1) the number of fast rotator/active stars excluded based on the cross match was only on the order of 1\% of the full APACHE sample; 2) the final ranking was dominated by parameters such as observability, the number of Gaia observations and the ongoing HADES spectroscopic monitoring. Therefore, although there may be a slight bias against fast rotators, we think the impact on the final distribution is very low. This is confirmed by the excellent agreement with the \textit{Kepler} survey period distribution, where a possible pre-selection against active stars/fast rotators is even more limited.

\section{Summary and discussion}
\label{sum}
We searched for photometric rotation periods of 247 M dwarfs in the mass range $0.15 - 0.70 M_\odot$  observed by the APACHE transit survey during five years of operations. 
When searching for sinusoidal signals in the data, we used GLS to calculate the frequency periodogram sampled on a uniform grid in frequency from 0.1 to 500 days.
We estimated the significance of the highest GLS peaks with a bootstrap analysis and we selected as candidate rotators all the objects with FAP $\leq$ 1\%, a semi-amplitude of the signal $\geq$ 2.5 mmag, and after a visual inspection.
The final sample consists of 107 M dwarfs with rotation periods in the range $0.52 - 191.82$ days, with a peak in distribution  at $\sim 30$ days.
We performed a simulation to test our survey sensitivity, and we quantified the loss of detection efficiency at long periods. This is an expected result due to the limited duration of the time series, the instrumental instability over five years and the intrinsic variation of stellar spots which we do not take into account in the period search.
In spite of this, we found that our period distribution is substantially unbiased in the period range $0.5-200$ days.
Looking at the rotation period-mass relation and excluding the fast rotators with $P < 10$ days, the period appears mass dependent, increasing with decreasing mass.
Moreover, for our sample of early M dwarfs, we found that the amplitude of variability is not correlated with the rotation period.

This catalogue of 107 M dwarfs with a measured rotation period bears potential for important contributions to the analysis of space-based photometric data from transit detection and characterization missions such as TESS, PLATO, CHEOPS, and ARIEL. In particular, in the presence of small-size transit candidates (of potential interest for atmospheric characterization too) our results might be useful in disentangling planetary signals (be they dynamical or atmospheric) from those related to stellar magnetic activity.

While the primary focus of this work is on the measurement of robust rotation periods of the APACHE sample of M dwarfs, it is nevertheless possible to comment on the statistics of the results obtained in the context of relevant aspects of the astrophysics of cool stars. 

We compared the rotation period of 23 APACHE targets with the rotation period derived spectroscopically by the HADES program. We see that for most stars there is a good correspondence between either the two rotation periods or one rotation period and the first harmonic of the other. 
Moreover, we compared our rotation period distribution with the distribution from 1570 rotating M dwarfs in the mass range $0.3-0.55$ $M_\odot$ observed by the \textit{Kepler} survey. 
The two distributions are in good agreement with the peak that lies around $\sim$ 30 days and a tail of fast rotators with P $<$ 10 days.
The comparison of our rotation periods to the spectroscopic periods from the literature and the comparison between our period distribution with the one from the \textit{Kepler} survey give support to the periods we detected, although we refer the reader to Sec.\,\ref{spec} and Sec.\,\ref{kepler} for the details.

We used the \emph{Gaia} DR2 to characterize the Galactic kinematics for our M dwarfs with measured rotation. We found that our sample has kinematics consistent with the  stellar number density distribution of the Milky Way in the Solar neighborhood.
Furthermore, we tested our kinematic characterization on the sub-sample of the late M dwarfs presented in \cite{irwin11}.
We can broadly group our rotating sample by their kinematics into the thin and thick disk galaxy population. This classification was used as a proxy for stellar age. While the thin disk sample spans the full range of periods and includes entirely the rapidly rotating objects, the thick disk sample is constitute only by slowly rotating stars.
Similarly to the thin disk sample, the stars with an intermediate kinematics classification span the full range of periods, but with no objects in the ultra fast ($P < 5$ days) rotating clump.
Given the mean old age of the thick disk ($\sim10$ Gyr), this suggests that the older objects in the sample are rotating more slowly. Based on a Spearman rank correlation test, that returned a value of the correlation coefficient of $0.22 \pm 0.03$,  we confirmed a (weak) correlation between the space velocity dispersion and the rotation period. 

Assuming that the velocity dispersion increases with age, as expected for older stellar population, the star's ages are increasing with rotation period. We investigated this aspect considering the velocity dispersion-age relation in the $V_Z$ velocity component, which is the most sensitive to age. 
For our most populated bin with period $ 10 < P < 70$ days, we found a mean age of $3.3 \pm 0.6 $ Gyr. We are not able to really constrain the ages of the other bins, due to the small size of our sample. Comparing this result with the one published by \cite{newton2016}, in which a mean age of $4.5^{+3.9}_{-2.3}$ Gyr was found for the period bin $ P > 70$ days, we confirm that late M dwarfs spin down to longer periods than early M dwarfs.

Despite the good agreement between the two works, there is a question still unresolved: the lack of fast rotators ($ P < 10$ days) in the APACHE/\textit{Kepler} sample.
This could be explained with an observational bias in the target selection with no ``young'' stars (age $< 1$ Gyr). In any case, we prioritized our targets mostly on observational constraints and taking into account the number of \emph{Gaia} transits, therefore our sample should not be astrophysically biased. Over an initial sample of $\sim$ 3000 M dwarfs only a handful of stars have been excluded as suspect young and over-active. This is more true for the \textit{Kepler} sample where it is difficult to imagine an observational bias based on the age/activity because of the lack of archive information for this type of stars due to their lower intrinsic luminosity.
On the other hand, we could assume an incorrect estimate of the mean age of late type M dwarf fast rotators or, symmetrically, a very different time scale in the rotation period evolution between early and late type stars. In this case, we assume that the late M dwarfs are maintaining rapid rotation for longer than their early counterparts or, symmetrically, the early M dwarfs spin down faster than their late counterparts. 
Indeed, the increase in the fraction of rapid rotators with decreasing stellar mass, with a particularly sharp increase in fast rotators at around $\sim0.3$ M$_\odot$ (which corresponds to the transition to fully convective stellar interiors) has recently been noticed by others (e.g., \citealt{gilhool18}, and references therein). The more natural explanation for this bimodality calls into question the effectiveness with which fully convective stars, late-type stars are capable of releasing angular momentum in comparison with earlier-type stars with radiative cores. 

\section*{Acknowledgements}
P.G. acknowledges financial support from the Italian Space Agency (ASI) under contract 2014-025-R.1.2015 to INAF;
M.D. acknowledges financial support from Progetto Premiale 2015 FRONTIERA funding scheme of the Italian Ministry of Education, University, and Research;
M.Pi. gratefully acknowledges the support from the European Union Seventh Framework Programme (FP7/2007-2013) under Grant Agreement No. 313014 (ETAEARTH);
D.B. acknowledges financial support from INAF and Agenzia Spaziale Italiana (ASI grant no. 2014-025-R.1.2015) for the 2016 Ph.D. fellowship programme of INAF;
EP acknowledges the financial support of the 2014 PhD fellowship programme of the INAF;
The Astronomical Observatory of the Autonomous Region of the Aosta Valley (OAVdA) is managed by the Fondazione Cl\'ement Fillietroz-ONLUS, which is supported by the Regional Government of the Aosta Valley, the Town Municipality of Nus and the ``Unit\'e des Communes vald\^otaines Mont-\'Emilius''. The authors thank ASI (through contracts I/037/08/0 and I/058/10/0) and Fondazione CRT for their support to the APACHE Project;
We acknowledge R. Smart, B. Bucciarelli and A. Spagna for interesting and useful discussions on various aspects of this work;
P.G. acknowledges Ilaria Carleo for unique conversations and contributions during the manuscript draft.



%
\bibliographystyle{mnras}
\bibliography{bibliogra} 




\begin{appendix}

\section{Tables}


\onecolumn
\footnotesize
\addtolength{\tabcolsep}{-0.7pt}
\begin{longtable}{rrrrrrrrrrr}

\caption{\label{tab:par2} Observational Properties for the APACHE stars.}\\
\hline
Name       &    RA         & Dec       & V                   & N points      & N nights     & $\Delta HJD$                           & $\sigma_t$ & $\Phi_{det}$ &  $\Phi_{det,h}$  & Rotator\\

           &     (deg)      & (deg)     & (mag)                 &   &   & (days)    & (mag) & (\%)  & (\%)  & Yes/No \\
\hline
\endfirsthead
\caption{Continued.} \\
\hline
Name       &    RA         & Dec       & V                   & N points      & N nights     & $\Delta HJD$                           & $\sigma_t$ &  $\Phi_{det}$ &  $\Phi_{det,h}$  & Rotator\\

           &     (deg)      & (deg)     & (mag)                 &   &   & (days)    & (mag) & (\%) & (\%) & Yes/No \\
\hline
\endhead
\hline
\endfoot
   J00156+7217 &     3.902023 &    72.283562 &  12.70 &    4798 &    108 &     814. &  0.0045 &      82 &    0 &  Y  \\ 
   J00166+3000 &     4.168521 &    30.016624 &  12.79 &    1772 &     39 &     481. &  0.0046 &      50 &    7 &  Y  \\ 
   J00207+5936 &     5.198862 &    59.604836 &  12.80 &    4935 &    138 &     798. &  0.0036 &      99 &    0 &  Y  \\ 
   J00234+7711 &     5.870607 &    77.189285 &  11.31 &    3718 &     99 &     618. &  0.0052 &      86 &    0 &  Y  \\ 
   J00374+5133 &     9.358173 &    51.552032 &  11.54 &    3168 &    103 &    1148. &  0.0043 &      94 &    0 &  Y  \\ 
   J01013+6121 &    15.333420 &    61.365776 &  10.92 &    4737 &     69 &     197. &  0.0083 &      61 &    0 &  Y  \\ 
   J01025+7140 &    15.633533 &    71.679871 &   9.98 &    4310 &     96 &    1538. &  0.0039 &     100 &    0 &  Y  \\ 
   J01056+2829 &    16.406815 &    28.492657 &  15.64 &    3159 &     82 &     800. &  0.0040 &      99 &    0 &  Y  \\
\end{longtable}

\twocolumn

\onecolumn
\footnotesize
\addtolength{\tabcolsep}{-0.7pt}
\begin{longtable}{rrrrrrrrrrrr}

\caption{\label{tab:par1} Stellar Parameters, Kinematics and Rotation Periods for the APACHE rotation candidates.} \\
\hline
Name       & B-V                    & Plx     & $T_{eff}$  & Mass & Radius    & $V_R$  & $V_{\phi}$ & $V_Z$ & Period & Amp   & Fap \\

           & (mag)                 & (mas) & (K) & ($M_\odot$) & ($R_\odot$)  & (km/s) & (km/s) & (km/s) & (Days)  & (mag) &  \\
\hline
\endfirsthead
\caption{Continued.} \\
\hline
Name       & B-V                    & Plx     & $T_{eff}$  & Mass & Radius    & $V_R$  & $V_{\phi}$ & $V_Z$ & Period & Amp   & Fap \\

           & (mag)                 & (mas) & (K) & ($M_\odot$) & ($R_\odot$)  & (km/s) & (km/s) & (km/s) & (Days)  & (mag) &  \\
\hline
\endhead
\hline
\endfoot
     J00156+7217 &   1.680  &    39.0409 $\pm$   0.0441 & $3848_{-120}^{+179}$ &   0.415 $\pm$   0.077 &   0.391 $\pm$   0.130 &     --- &     --- &   --- &    46.8713 &  0.0039 &  0.0001  \\
   J00166+3000 &   1.492  &    24.0628 $\pm$   0.1735 & $3685_{-221}^{+595}$ &   0.250 $\pm$   0.094 &   0.252 $\pm$   0.122 &    27.9 &   226.2 &   5.7 &    13.9112 &  0.0028 &  0.0017  \\
   J00207+5936 &   1.433  &    35.0861 $\pm$   0.0436 & $3794_{-79}^{+151}$  &   0.380 $\pm$   0.053 &   0.341 $\pm$   0.043 &   -10.0 &   260.2 &  -7.9 &    22.0334 &  0.0037 &  0.0000  \\
   J00234+7711 &   1.591  &    52.2518 $\pm$   0.0209 & $3805_{-105}^{+146}$ &   0.410 $\pm$   0.073 &   0.384 $\pm$   0.124 &  -110.8 &   228.5 &   3.8 &    18.3830 &  0.0104 &  0.0020  \\
   J00374+5133 &   1.425  &    30.9169 $\pm$   0.0467 & $3900_{-100}^{+128}$ &   0.470 $\pm$   0.064 &   0.437 $\pm$   0.109 &  -105.8 &   240.1 &  17.8 &    15.5063 &  0.0094 &  0.0000  \\
   J01013+6121 &   1.595  &    95.0899 $\pm$   0.0372 & $3743_{-164}^{+119}$ &   0.355 $\pm$   0.069 &   0.337 $\pm$   0.072 &     6.8 &   239.4 & -32.1 &    75.3565 &  0.0068 &  0.0000  \\
   J01025+7140 &   1.409  &   121.4237 $\pm$   0.0339 & $3786_{-221}^{+177}$ &   0.330 $\pm$   0.053 &   0.305 $\pm$   0.038 &    46.2 &   210.0 &  -3.7 &    49.4642 &  0.0039 &  0.0001  \\
\end{longtable}
\twocolumn


\end{appendix}

\bsp	
\label{lastpage}
\end{document}